\documentclass[10pt, prb, aps, twocolumn, showpacs, citeautoscript, floatfix, reprint, amsmath, amssymb, notitlepage, superscriptaddress]{revtex4-1}

\usepackage{booktabs}
\usepackage[utf8]{inputenc}
\usepackage[T1]{fontenc}
\usepackage[english]{babel}
\usepackage{graphicx}
\usepackage{rotating}
\usepackage{mathtools}
\usepackage{amsfonts}
\usepackage{dcolumn} % align table columns on decimal point
\usepackage{bm} % bold math
\usepackage{xcolor}
\usepackage{latexsym}
\usepackage{wasysym}
\usepackage[colorlinks, citecolor=blue, linkcolor=blue, urlcolor=blue]{hyperref}
\usepackage[nomain, acronym, shortcuts]{glossaries}
\usepackage{tikz}
\usepackage{environ}
\usepackage{csquotes}
\glsdisablehyper
\usetikzlibrary{external}
\bibpunct{[}{]}{,}{n}{}{}

\begin{document}

\title{Weakly Interacting Topological Insulators: Quantum Criticality and \\ Renormalization Group Approach}

\author{Wei Chen}

\affiliation{Department of Physics, PUC-Rio, Rio de Janeiro, Brazil}

\date{\today}

\begin{abstract}

For $D$-dimensional weakly interacting topological insulators in certain symmetry classes, the topological invariant can be calculated from a $D$- or $(D+1)$-dimensional integration over a certain curvature function that is expressed in terms of single-particle Green's functions. Based on the divergence of curvature function at the topological phase transition, we demonstrate how a renormalization group approach circumvents these integrations and reduces the necessary calculation to that for the Green's function alone, rendering a numerically efficient tool to identify topological phase transitions in a large parameter space. The method further unveils a number of statistical aspects related to the quantum criticality in weakly interacting topological insulators, including correlation function, critical exponents, and scaling laws, that can be used to characterize the topological phase transitions driven by either interacting or noninteracting parameters. We use 1D class BDI and 2D class A Dirac models with electron-electron and electron-phonon interactions to demonstrate these principles, and find that interactions may change the critical exponents of the topological insulators. 

\end{abstract}


%64.60.ae Renormalization-group theory

%64.60.F- Equilibrium properties near critical points, critical exponents

%73.20.-r Electron states at surfaces and interfaces

\maketitle

\section{Introduction}

The effect of many-body interaction on the topological properties of materials has been a fascinating subject in the research of topological insulators (TIs). From the symmetry perspective, the noninteracting TIs have been well understood within the context of symmetry classification\cite{Schnyder08,Kitaev09,Chiu16}, and a significant amount of effort has been made to generalize the symmetry classification scheme to incorporate many-body interactions\cite{Morimoto15,Queiroz16,Song17,Wang17}. On the other hand, through investigating concrete models, interactions have been shown to lead to novel and intriguing phenomena, such as the notion of topological Mott insulators\cite{Raghu08}, fractional topological insulators\cite{Levin09,Maciejko10,Maciejko15}, topological Kondo insulators\cite{Dzero10}, dynamical axion field\cite{Li10}, fractional Chern insulators (FCIs)\cite{Neupert11,Regnault11}, and first-order topological phase transitions\cite{Amaricci15,Amaricci16,Imriska16}, among many others.

Limiting our discussion to weakly interacting TIs, a suitable framework to address the effect of interactions is the Green's function formalism\cite{So85,Niu85,Ishikawa86,Volovik88,Volovik03,Qi08,Wang10,Gurarie11,
Essin11,Wang12,Wang12_2,Wang13,Grandi15}.
Within this formalism, the quantity that characterizes the topology of the system, namely the topological invariant, can be expressed in terms of the single-particle Green's function. In particular, in certain symmetry classes, the topological invariant of a $D$-dimensional TI takes the form of a $D$-dimensional momentum space integral or a $(D+1)$-dimensional momentum-frequency space integral over a certain combination of the single-particle Green's function and its derivatives. For the rest of the article, the integrand in this $D$- or $(D+1)$-dimensional integral is referred to as the curvature function, synonymous to the integration of the local curvature of a closed string counts the number of knots it contains. Depending on the symmetry class, the curvature function takes different forms. The advantage of this Green's function formalism is that the many-body effects, such as disorder or correlations, can be conveniently incorporated in a perturbative manner, provided the system remains in the weak coupling regime and is continuously connected to a noninteracting topological phase\cite{He16_2}.

To investigate the many-body effect for these particular symmetry classes using the Green's function formalism, however, appears to be a tedious task in practice. This is because within the framework of perturbation theory, the Green's function itself already contains at least a $D$-dimensional momentum space integration for the self-energy, making it totally a $2D$- or $(2D+1)$-dimensional integration problem. This is obviously a very costly calculation, especially if one's interest is to search for the topological phase transitions in a large parameter space for TIs defined at some high spatial dimension.

A generic feature near topological phase transitions seems to indicate an algorithm that can circumvent these integrations. In noninteracting systems, in which the curvature function is calculated from Berry connection, Berry curvature\cite{Chen16}, or Pfaffian of the time-reversal operator\cite{Chen16_2}, one observes that the curvature function generally diverges at a certain high symmetry point (HSP) in the momentum space as the system approaches the critical point, and the divergence changes sign across the critical point. This divergence is essentially due to the closing of the bulk gap at the HSP. The topological phase transition that features this divergence is said to be of second-order. Recently, it is revealed that FCIs near interaction-driven topological phase transitions also manifest such a divergence, in which the curvature function is a many-body Berry curvature calculated within the twisted boundary condition\cite{Kourtis17}. Based on this divergence, a renormalization group approach (RG) has been proposed\cite{Chen16,Chen16_2,Kourtis17}, which is a iterative procedure to find the trajectory (RG flow) in the parameter space along which the divergence is reduced, from which the topological phase transitions can be identified. In this article, we will demonstrate that, at least for the Dirac models and the interactions we examined, the curvature function features such a divergence and hence the RG approach is applicable. A significance of this RG approach is that it only requires to calculate the curvature function at few points in the momentum or momentum-frequency space, and consequently the $D$- or $(D+1)$-dimensional integration of the curvature function is avoided.

Besides demonstrating the RG approach as a numerically convenient tool, another purpose of this article is to introduce the following statistical aspects into weakly interacting TIs based on the divergence of the curvature function\cite{Chen17}, using 1D class BDI and 2D class A Dirac models in the presence of electron-electron and electron-phonon interaction as examples. The first is a real space or spacetime correlation function as the Fourier transform of the curvature function, which takes the form of a product of real space or spacetime Green's function propagated over several segments. Remarkably, at the critical point, the correlation length or correlation time diverges due to the divergence of the curvature function, signifying the scale invariance. The critical exponents of the correlation length and correlation time, as well as that of the curvature function at the HSP, are shown to be constrained by a scaling law due to conservation of the topological invariant. In addition, through investigating concrete models, we unveils that interaction may affect the critical exponents of the system.

The article is structured in the following manner. In Sec.~II, we introduce the general formalism of Green's function and the perturbative approach to Dirac models under weak electron-electron and electron-phonon interactions. The divergence of curvature function and the RG approach are then introduced on a phenomenological level. In Sec.~III, we discuss 1D class BDI models with interactions, including the specific form of topological invariant and the accompanied correlation function. A concrete model is then used to demonstrate the RG approach and to extract the critical exponents. In Sec.~IV, we discuss 2D class A models, for both of frequency-dependent and frequency-independent self-energies. We introduce the correlation function for both cases, and discuss how the interactions influence the critical exponents. Sec.~V summarizes the results.

\section{Generic critical behavior investigated by perturbative treatment\label{sec:generic_critical_perturbative}}

\subsection{Green's function formalism for Dirac models with interactions\label{sec:Green_fn_Dirac_interaction}}

Our goal is to use Green's function formalism to study the $2\times 2$ Dirac models under the influence of many-body interactions, and also to use the Green's function to calculate the many-body curvature function introduced in the next section, so here we briefly outline the plan of attack. Starting from a noninteracting spinless $2\times 2$ Dirac Hamiltonian
\begin{eqnarray}
&&{\cal H}_{0}=\sum_{\bf k}
\left(
\begin{array}{cc}
c_{A{\bf k}}^{\dag} & c_{B{\bf k}}^{\dag} \\
\end{array}
\right)H_{0}({\bf k})
\left(
\begin{array}{c}
c_{A{\bf k}} \\
c_{B{\bf k}}
\end{array}
\right)\;,
\nonumber \\
&&H_{0}({\bf k})=d_{0}({\bf k})\sigma_{0}+d_{1}({\bf k})\sigma_{1}+d_{2}({\bf k})\sigma_{2}+d_{3}({\bf k})\sigma_{3}\;.
\label{general_2by2_Dirac_Hamiltonian}
\end{eqnarray}
which is the suitable minimal model for 1D class BDI and 2D class A systems. Depending on the symmetry class, not every $d_{i}$ is present, and also the eveness and oddness of each $d_{i}$ depends on the symmetry class. The two degrees of freedoms will be referred to as $A$ and $B$ sublattices, although in general it may represent some other pseudospin degrees of freedom. The $d_{0}$ component may be set to zero at noninteracting level, but interactions can induce $d_{0}$, as explain later. Here $c_{I{\bf k}}$ denotes the spinless fermion operator that satisfies the anticommutation relation $\left\{c_{I{\bf k}},c_{I^{\prime}{\bf k}^{\prime}}^{\dag}\right\}=\delta_{II^{\prime}}\delta_{{\bf k}{\bf k}^{\prime}}$ with sublattice index $I=\left\{A,B\right\}$.

In the presence of a weak interaction written in the second-quantized form ${\cal H}_{int}$, our strategy is to calculate the effect of interaction perturbatively using Matsubara Green's function
\begin{eqnarray}
G({\bf k},\tau)=\left(
\begin{array}{ll}
G_{AA}({\bf k},\tau) & G_{AB}({\bf k},\tau) \\
G_{BA}({\bf k},\tau) & G_{BB}({\bf k},\tau)
\end{array}
\right)\;,
\end{eqnarray}
with the matrix elements defined by 
\begin{eqnarray}
G_{IJ}({\bf k},\tau)=-\langle T_{\tau}c_{I{\bf k}}(\tau)c_{J{\bf k}}^{\dag}(0)\rangle\;.
\end{eqnarray}
where $T_{\tau}$ is the time ordering. The full Green's function with discrete frequency $i\omega_{n}$ can be obtained from Dyson's equation
\begin{eqnarray}
G&=&G_{0}+G_{0}\Sigma G=G_{0}+G_{0}\Sigma G_{0}+G_{0}\Sigma G_{0}\Sigma G_{0}+...
\nonumber \\
&=&\left(G_{0}^{-1}-\Sigma\right)^{-1}=\left(i\omega_{n}-H_{0}-\Sigma\right)^{-1},
\label{Dyson_equation}
\end{eqnarray}
where $\Sigma$ is the self-energy. The interacting part of Dyson's equation in Eq.~(\ref{Dyson_equation}) is
\begin{eqnarray}
&&\left(G_{0}\Sigma G\right)_{IJ}=-\sum_{n=1}^{\infty}(-1)^{n}\int_{0}^{\beta}d\tau_{1}\int_{0}^{\beta}d\tau_{2}...
\int_{0}^{\beta}d\tau_{n}
\nonumber \\
&&\times\langle T_{\tau} c_{I{\bf k}}(\tau){\cal H}_{int}(\tau_{1}){\cal H}_{int}(\tau_{2})...{\cal H}_{int}(\tau_{n})c_{J{\bf k}}^{\dag}(0)\rangle,
\end{eqnarray}
that evaluates different, connected diagrams. In this article we restrict our calculation to one-loop level. 

%{\cblue (Note that one-loop level is $n=1$ for density-density interaction and $n=2$ for electron-phonon interaction in the above equation. Should mention this?)}

To be specific, we examine two kinds of interactions. The first is the short range interaction between spinless fermions\cite{Wen10,Grushin13,Jia13,Daghofer14,Guo14,Luo15}. In particular, we consider the density-density interaction between two sublattices that takes the form
\begin{eqnarray}
{\cal H}_{e-e}=\sum_{\bf kk^{\prime}q}V_{\bf q}c_{A{\bf k+q}}^{\dag}c_{B{\bf k^{\prime}-q}}^{\dag}c_{B{\bf k}^{\prime}}c_{A{\bf k}}\;.
\label{general_density_density_interaction}
\end{eqnarray}
The precise form of $V_{\bf q}$ will be discussed in later sections. The second is the electron-phonon interaction\cite{Zhu12,Li12,Kim12,Huang12,Parente13,DasSarma13,Li13,Luo13,Sobota14,
Howard14,Gupta14,Glinka15,Glinka15_2,Zhao15,Sharafeev17,Tamtogl17,Heid17}, which has been a realistic issue, for instance on the transport of surface states. To examine the perturbative approach, we consider a polar coupling between spinless fermions and a longitudinal optical phonon mode, described by the Fr\"{o}hlich Hamiltonian\cite{Mahan00}
\begin{eqnarray}
{\cal H}_{e-ph}&=&\sum_{\bf kq}M_{\bf q}\left(c_{A{\bf k+q}}^{\dag}c_{A{\bf k}}+c_{B{\bf k+q}}^{\dag}c_{B{\bf k}}\right)A_{\bf q}\;,
\nonumber \\
A_{\bf q}&=&a_{\bf q}+a_{\bf -q}^{\dag}\;,
\nonumber \\
M_{\bf q}&=&\frac{u\,\omega_{0}}{q}\;.
\label{general_electron_phonon_interaction}
\end{eqnarray}
Here $a_{\bf q}$ is the phonon annihilation operator, $\omega_{0}$ is the optical phonon frequency, and $u$ is a phenomenological prefactor that takes care of the dielectric constant. We will treat all the variables as dimensionless for simplicity. 

%{\cblue (Well, if I discuss the symmetry of density-density interaction then it seems like I also have to describe the symmetry of the electron-phonon interaction)}

The calculation of self-energy for these two types of interactions is detailed in Appendix \ref{appendix:self_energy}. After the Dyson's equation is solved, we then treat the discrete Matsubara frequency continuously $i\omega_{n}\rightarrow i\omega$ to obtain the Green's function with continuous frequency variable for the sake of calculating the topological invariant, as detailed in the following sections. Taking into account how the self-energy enters the Green's function, the full Green's function in momentum-frequency space takes the form
\begin{eqnarray}
G({\bf k},i\omega)
&=&\frac{1}{(i\omega+d_{0}^{\prime})^{2}-d^{\prime 2}}
\nonumber \\
&&\times\left(
\begin{array}{cc}
i\omega+d_{0}^{\prime}+d_{3}^{\prime} & d_{1}^{\prime}-id_{2}^{\prime} \\
d_{1}^{\prime}+id_{2}^{\prime} & i\omega+d_{0}^{\prime}-d_{3}^{\prime}
\end{array}
\right)\;,
\nonumber \\
G^{-1}({\bf k},i\omega)&=&\left(
\begin{array}{cc}
i\omega+d_{0}^{\prime}-d_{3}^{\prime} & -d_{1}^{\prime}+id_{2}^{\prime} \\
-d_{1}^{\prime}-id_{2}^{\prime} & i\omega+d_{0}^{\prime}+d_{3}^{\prime}
\end{array}
\right)\;,
\label{2D_G_dprime}
\end{eqnarray}
where $d^{\prime}=\sqrt{d_{1}^{\prime 2}+d_{2}^{\prime 2}+d_{3}^{\prime 2}}$, and the self-energy-renormalized ${\bf d}^{\prime}$-vector is
\begin{eqnarray}
&&d_{1}^{\prime}=d_{1}+{\rm Re}\Sigma_{AB}\;,\;\;\;d_{2}^{\prime}=d_{2}-{\rm Im}\Sigma_{AB}\;,
\nonumber \\
&&d_{3}^{\prime}=d_{3}+\frac{\Sigma_{AA}-\Sigma_{BB}}{2}\;,
\;\;\;d_{0}^{\prime}=\frac{-\Sigma_{AA}-\Sigma_{BB}}{2}\;.
\label{renormalized_d_vector_mu}
\end{eqnarray}
Notice that because of the self-energy $\Sigma_{IJ}=\Sigma_{IJ}({\bf k},i\omega)$, each $d_{i}^{\prime}=d_{i}^{\prime}({\bf k},i\omega)$ is generally a function of both momentum and frequency. In 1D class BDI models, the $i\omega$ is eventually analytically continued to $i\omega\rightarrow E$ and then taken as a real number, whereas in 2D class A models the $i\omega$ remains a imaginary number that is to be integrated out.

%The frequency-dependence also implies that in the presence of interactions, the quasiparticle spectral function is in general no longer a $\delta$-function, hence the notion of band gap becomes rather ambiguous, and so is to identify topological phase transitions from gap-closing. Rather, the transitions may be identified from the divergence of the curvature function, as introduced in the next section.

\subsection{Divergence of curvature function\label{sec:criticality}}

In noninteracting TIs, the topological phase transition can be identified from the closing of the bulk gap. However, in interacting systems, the single-particle spectral function is in general not a $\delta$-function but spreads out over a range of frequency, so the bulk gap is strictly speaking a rather ambiguous notion. In this case, the Green's function formalism offers a way to identify the topological phase transition without explicitly invoking the notion of bulk gap. In this article, we examine the Green's function formalism and discuss the quantum criticality near the topological phase transitions within the context of the divergence of the curvature function, as introduced below.

In the Green's function formalism, the integer-valued topological invariant ${\cal C}={\cal C}({\bf M})$ is formulated in terms of single-particle Green's functions, and depends on a set of tuning parameters ${\bf M}=(M_{1},M_{2}...M_{D_{M}})$ that form a $D_{M}$-dimensional parameter space. Each $M_{i}$ is either a noninteracting or interacting parameter in the Hamiltonian. The topological invariant for the two classes under consideration takes the form of a $\tilde{D}$-dimensional integration over momentum space or momentum-frequency space
\begin{eqnarray}
{\cal C}&=&\int d^{\tilde D}{\bf K}\,F({\bf K},{\bf M})\;,
\nonumber \\
\tilde{D}&=&1\;,\;{\bf K}=k\;{\rm for\;1D\;class\;BDI}\;,
\nonumber \\
\tilde{D}&=&3\;,\;{\bf K}=(\omega,k_{x},k_{y})\;{\rm for\;2D\;class\;A}.
\label{C_FkM_class_BDI_class_A}
\end{eqnarray}
The function $F({\bf K},{\bf M})$ is referred to as the curvature function, and is expressed in terms of the single-particle Green's function calculated perturbatively in Sec.~\ref{sec:Green_fn_Dirac_interaction}. We assume that there is a certain HSP in the momentum or momentum-frequency space, denoted by ${\bf K}_{0}$ and satisfying ${\bf K}_{0}=-{\bf K}_{0}$, around which the curvature function is an even function 
\begin{eqnarray}
F({\bf K}+\delta{\bf K},{\bf M})=F({\bf K}_{0}-\delta{\bf K},{\bf M})\;
\end{eqnarray}
due to certain symmetries, such as inversion symmetry, where $\delta{\bf K}$ is a small displacement.

%The curvature function offers an opportunity to characterize the quantum criticality near the topological phase transition driven by either interacting or noninteracting parameters. 

\begin{figure}[ht]
\begin{center}
\includegraphics[clip=true,width=0.99\columnwidth]{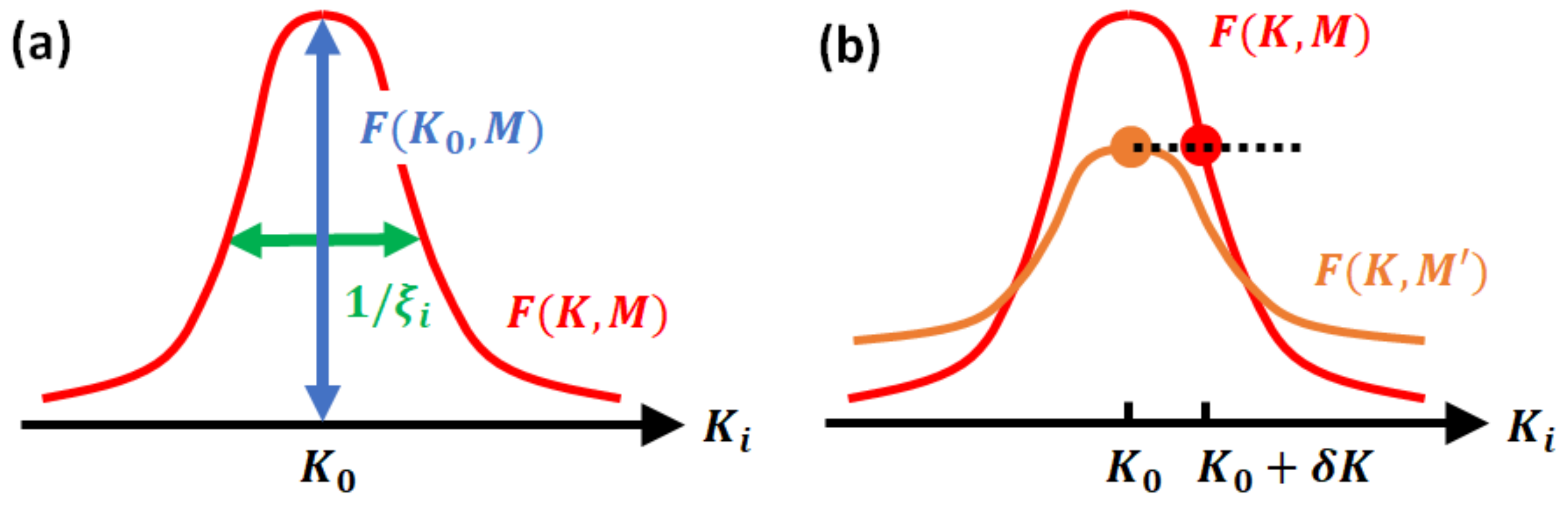}
\caption{ (a) Schematics of the Ornstein-Zernike form of the curvature function in the vicinity of a HSP ${\bf K}_{0}$ along certain direction $K_{i}$ in the momentum or momentum-frequency space. The constant volume underneath the Lorentzian peak leads to the scaling law in Eq.~(\ref{scaling_law}) that constrains the critical exponent of $F({\bf K}_{0},{\bf M})$ and that of $\xi_{i}$. (b) Schematics of the RG procedure, which demands $F({\bf K}_{0}+\delta{\bf K},{\bf M})$ (red dot) to be equal to $F({\bf K}_{0},{\bf M}^{\prime})$ (orange dot), as indicated by the dashed line. The divergence of the curvature function at ${\bf K}_{0}$ is gradually reduced under this procedure whereas the topological invariant remains unchanged, hence the system gradually flows away from the critical point. } 
\label{fig:Fk0_xi_RG_schematics}
\end{center}
\end{figure}

At least for the models investigated in this article, we observe the following critical behavior for the curvature function. Firstly, the curvature function around the HSPs takes the Ornstein-Zernike form in all $\tilde{D}$-directions in the momentum or momentum-frequency space
\begin{eqnarray}
F({\bf K}_{0}+\delta{\bf K},{\bf M})=\frac{F({\bf K}_{0},{\bf M})}{\prod_{i=1}^{\tilde{D}}\left(1+\xi_{i}^{2}\delta K_{i}^{2}\right)}\;,
\label{Ornstein_Zernike}
\end{eqnarray}
where $\xi_{i}$ is a length or time scale, as indicated schematically in Fig.~\ref{fig:Fk0_xi_RG_schematics}. Secondly, as approaching the critical point ${\bf M}\rightarrow{\bf M}_{c}$, the extremum $F({\bf K}_{0},{\bf M})$ diverges, and the time or length scale $\xi_{i}$ diverges or saturates to a constant
\begin{eqnarray}
&&\lim_{{\bf M}\rightarrow{\bf M}_{c}^{+}}F({\bf K}_{0},{\bf M})=-\lim_{{\bf M}\rightarrow{\bf M}_{c}^{-}}F({\bf K}_{0},{\bf M})=\pm\infty\;,
\nonumber \\
&&\lim_{{\bf M}\rightarrow{\bf M}_{c}}\xi_{i}=\infty\;{\rm or}\;{\rm const.},
\label{Fk0_xi_divergence}
\end{eqnarray}
where ${\bf M}_{c}^{+}$ and ${\bf M}_{c}^{-}$ denote the two sides of the phase boundary in the parameter space. In short terms, the Lorentzian shape of Eq.~(\ref{Ornstein_Zernike}) gradually narrows as ${\bf M}\rightarrow{\bf M}_{c}$, and flips sign abruptly as ${\bf M}$ crosses ${\bf M}_{c}$. The critical exponents then follows
\begin{eqnarray}
F({\bf K}_{0},{\bf M})\propto |{\bf M}-{\bf M}_{c}|^{-\gamma}\;,\;\;\;\xi_{i}\propto |{\bf M}-{\bf M}_{c}|^{-\nu_{i}}\;.
\label{Fk0_xi_critical_exponent}
\end{eqnarray}
The numerator $F({\bf K}_{0},{\bf M})$ in the Ornstein-Zernike form of Eq.~(\ref{Ornstein_Zernike}) is assigned to the exponent of the susceptibility $\gamma$, and the length or time scale $\xi_{i}$ is assigned to the exponent of the correlation length $\nu_{i}$, for the reason that will become clear later.  

The critical exponents $\gamma$ and $\nu_{i}$ are not independent, but constrained by a scaling law due to the conservation of topological invariant as ${\bf M}\rightarrow {\bf M}_{c}$. This can be seen by considering the integration of the Ornstein-Zernike form of Eq.~(\ref{Ornstein_Zernike}) over the width of the Lorentzian peak, which contributes to a fraction of the topological invariant that remains constant 
\begin{eqnarray}
{\cal C}_{div}&=&F({\bf K}_{0},{\bf M})\prod_{i=1}^{\tilde D}\left(\int_{-\xi_{i}^{-1}}^{\xi_{i}^{-1}}\frac{dK_{i}}{1+\xi_{i}^{2}K_{i}^{2}}\right)
\nonumber \\
&=&\frac{F({\bf K}_{0},{\bf M})}{\prod_{i=1}^{\tilde D}\xi_{i}}\times{\cal O}(1)={\rm Const}.
\end{eqnarray}
A scaling law then follows after applying Eq.~(\ref{Fk0_xi_critical_exponent}), 
\begin{eqnarray}
\gamma=\sum_{i=1}^{\tilde D}\nu_{i}\;,
\label{scaling_law}
\end{eqnarray}
which constraints the critical exponents\cite{Chen17}.

Previous investigations\cite{Chen17,Kourtis17} seem to suggest a general principle of introducing the correlation function that characterizes the topological phase transitions, which is found to also apply to weakly interacting systems: if the topological invariant is an integral of a certain curvature function, then the Fourier transform of the curvature function represents a correlation function
\begin{eqnarray}
\lambda_{\bf R}&=&\int d^{\tilde D}{\bf K}\,e^{i{\bf K}\cdot{\bf R}}F({\bf K},{\bf M})\;,
\nonumber \\
{\bf R}&=&
\left\{
\begin{array}{l}
r\;{\rm for\;1D\;class\;BDI}\;,\\
(-t,r_{x},r_{y})\;{\rm for\;2D\;class\;A}\;.
\end{array}
\right.
\label{general_Fourier_trans_lambdaR}
\end{eqnarray}
Here for 2D class A one has $e^{i{\bf K}\cdot{\bf R}}=e^{ik_{x}r_{x}+ik_{y}r_{y}-i\omega t}$ (defining $R_{0}=t$ or $-t$ does not make a difference since the curvature function is even in $\omega$, but we follow the usual convention of Fourier transform). Together with the Ornstein-Zernike form of the curvature function in Eq.~(\ref{Ornstein_Zernike}), this implies the correlation function decays exponentially in all ${\tilde D}$-directions with a correlation length or correlation time $\xi_{i}$. In contrast to previous investigations that consider certain kinds of two-point correlation of the field operator\cite{Ringel11,You14,Toldin15}, the $\lambda_{\bf R}$ in Eq.~(\ref{general_Fourier_trans_lambdaR}) is generally a product of real space or spacetime Green's functions propagated over several segments, as we shall see in the following sections. This Fourier transform principle is motivated by the progress in the theory of charge polarization and theory of orbital magnetization, which are real space representations of the topological order in noninteracting systems, and the correlation function therein measures a certain overlap of Wannier functions\cite{Chen17}. We should also emphasize that the correlation functions are nonzero in both topologically trivial and nontrivial phases, hence a more general measure for the quantum criticality than the edge state, as the later only manifests in the topologically nontrivial phase.

\subsection{Renormalization group approach\label{sec:RG}}

The discussion in Sec.~\ref{sec:criticality} motivates us to apply a recently developed RG scheme\cite{Chen16,Chen16_2} to identify the topological phase transitions in the $D_{M}$-dimensional parameter space, which renders a very convenient tool that significantly reduces the numerical effort to solve the many-body problem at hand. The RG procedure is based on the divergence of the curvature function described by Eqs.~(\ref{Ornstein_Zernike}) and (\ref{Fk0_xi_divergence}). The procedure demands that at a given ${\bf M}$, one solves for the new ${\bf M}^{\prime}$ that satisfies
\begin{eqnarray}
F({\bf K}_{0}+\delta{\bf K},{\bf M})=F({\bf K}_{0},{\bf M}^{\prime})\;,
\label{RG_procedure_general}
\end{eqnarray}
where ${\bf K}_{0}$ is an HSP, and $\delta{\bf K}$ is a small deviation away from it along a scaling direction, as indicated schematically in Fig.~\ref{fig:Fk0_xi_RG_schematics}. Iteratively solving for the mapping ${\bf M}\rightarrow{\bf M}^{\prime}$ yields an RG flow in the parameter space along which the divergence of $F({\bf K}_{0},{\bf M})$ is reduced but the topological invariant remains unchanged, in a way analogous to stretching a messy string until the number of knots becomes obvious, from which the topological phase transitions can be identified. To obtain the RG flow for each tuning parameter $M_{i}$, we use Eq.~(\ref{RG_procedure_general}) but fix all other tuning parameters unchanged
\begin{eqnarray}
&&F({\bf K}_{0}+\delta{\bf K},(M_{1},M_{2},...M_{i}...M_{d_{M}}))
\nonumber \\
&&=F({\bf K}_{0},(M_{1},M_{2},...M_{i}^{\prime}...M_{d_{M}}))\;.
\label{scaling_procedure_each_Mi}
\end{eqnarray}
Denoting $M_{i}^{\prime}-M_{i}=dM_{i}$ and $dl=|\delta{\bf K}|^{2}$, expanding Eq.~(\ref{scaling_procedure_each_Mi}) to leading order yields the generic RG equation
\begin{eqnarray}
\frac{dM_{i}}{dl}=\frac{1}{2}\frac{(\delta {\bf K}\cdot{\boldsymbol\nabla}_{\bf K})^{2}F({\bf K},{\bf M})|_{{\bf K}={\bf K}_{0}}}{\partial_{M_{i}}F({\bf K}_{0},{\bf M})}\;.
\label{generic_RG_eq}
\end{eqnarray}
For interacting systems, often times the numerical calculation is performed on discrete mash points, in which case Eq.~(\ref{generic_RG_eq}) may be approximated by 
\begin{eqnarray}
\frac{dM_{i}}{dl}=\frac{\Delta M_{i}}{|\Delta{\bf K}|^{2}}\frac{F({\bf K}_{0}+\Delta{\bf K},{\bf M})-F({\bf K}_{0},{\bf M})}{F({\bf K}_{0},{\bf M}+\Delta{\bf M}_{i})-F({\bf K}_{0},{\bf M})}\;,
\label{generic_RG_eq_discrete}
\end{eqnarray}
where $\Delta {\bf K}$ is the grid spacing along the scaling direction in the momentum or momentum-frequency space, $\Delta M_{i}$ is the grid spacing along ${\hat{\bf M}}_{i}$ direction in the parameter space, and $\Delta {\bf M}_{i}=\Delta M_{i}{\hat{\bf M}}_{i}$. Equation (\ref{generic_RG_eq_discrete}) demonstrates the advantage of this RG scheme, namely at a given ${\bf M}$, one only needs to calculate the curvature function at three mesh points $F({\bf K}_{0},{\bf M})$, $F({\bf K}_{0}+\Delta{\bf K},{\bf M})$, and $F({\bf K}_{0},{\bf M}+\Delta{\bf M}_{i})$ to calculate the RG flow $dM_{i}/dl$ without explicitly performing the $\tilde{D}$-dimensional integration in Eq.~(\ref{C_FkM_class_BDI_class_A}). In addition, to extract the critical exponents in Eq.~(\ref{Fk0_xi_critical_exponent}) also requires no more than these mesh points, provided $\Delta{\bf K}$ is along the direction reciprocal to $\xi_{i}$. For an arbitrarily given model, one should apply RG to all HSPs since the curvature function may diverge at any of them, but in this article we focus on the HSP where the noninteracting model is known to close the bulk gap. We remark that this RG scheme is applicable even if the Green's function (and subsequently the curvature function) is calculated by means other than the perturbative treatment in Sec.~\ref{sec:Green_fn_Dirac_interaction}, for instance quantum Monte Carlo (QMC) or exact diagonalization. Because it renormalizes the curvature function, the RG scheme is hereafter referred to as the curvature renormalization group (CRG) approach.

In this CRG scheme, the critical point and fixed point are identified from the RG flow from the flow rate and the direction of the flow, summarized below:
\begin{eqnarray}
{\rm Critical\;point}:&&\left|\frac{d{\bf M}}{dl}\right|\rightarrow\infty,\;{\rm flow\;directs\;away},
\nonumber \\
{\rm Stable\;fixed\;point}:&&\left|\frac{d{\bf M}}{dl}\right|\rightarrow 0,\;{\rm flow\;directs\;into},
\nonumber \\
{\rm Unstable\;fixed\;point}:&&\left|\frac{d{\bf M}}{dl}\right|\rightarrow 0,\;{\rm flow\;directs\;away}.
\nonumber \\
\label{identifying_Mc_Mf}
\end{eqnarray}
The critical point ${\bf M}_{c}$ form a $(D_{M}-1)$-dimensional surface in the $D_{M}$-dimensional parameter space. The flow rate diverges at ${\bf M}_{c}$ because the numerator in Eq.~(\ref{generic_RG_eq}) diverges, as can be seen from the Ornstein-Zernike form in Eq.~(\ref{Ornstein_Zernike}) and the critical behavior in Eq.~(\ref{Fk0_xi_divergence}). The flow rate vanishes at the fixed point ${\bf M}_{f}$ because the numerator in Eq.~(\ref{generic_RG_eq}) vanishes, which corresponds to a vanishing correlation length or correlation time $\xi_{i}$. The divergence of $\xi_{i}$ at ${\bf M}_{c}$ and vanish of it at ${\bf M}_{f}$ is the interpretation of scale invariance in this CRG scheme,  since the amplitude of the correlation does not depend on the distance ${\bf R}$.

\section{1D class BDI with interactions\label{sec:1D_class_BDI}}

\subsection{Topological invariant in the presence of interactions\label{sec:Green_function_1D}}

We proceed to use 1D class BDI spinless fermionic systems to demonstrate the principles in Sec.~\ref{sec:generic_critical_perturbative}. The appropriate minimal model is that in Eq.~(\ref{general_2by2_Dirac_Hamiltonian}) with 
\begin{eqnarray}
d_{0,1}(k)=d_{0,1}(-k)\;,\;\;\;d_{2}(k)=-d_{2}(-k)\;.
\label{BDI_1D_d_vector}
\end{eqnarray}
That is, the noninteracting Dirac Hamiltonian only contains the $d_{0,1}$ components that are even in $k$, and the $d_{2}$ component that is odd in $k$. The $d_{3}$ component is absent. In the presence of interactions, the ${\bf d}$-vector is changed to the ${\bf d}^{\prime}$-vector according to Eq.~(\ref{renormalized_d_vector_mu}).

We adopt the topological invariant expressed in terms of the full Green's function\cite{Essin11}
\begin{eqnarray}
{\cal C}=\int_{0}^{2\pi} \frac{dk}{4\pi i} \;{\rm Tr}\left(\sigma_{3}G^{-1}\partial_{k}G\right)|_{i\omega=-d_{0}^{\prime}}\;,
\label{topo_invariant}
\end{eqnarray}
evaluated at $i\omega=-d_{0}^{\prime}$, where $-d_{0}^{\prime}$ can be viewed as the interaction-induced chemical potential. In the case of density-density interaction in which the self-energy is frequency-independent, the requirement of $i\omega=-d_{0}^{\prime}$ physically means that the entire spectrum is shifted by $-d_{0}^{\prime}$ and hence the spectrum is particle-hole symmetric with respect to $-d_{0}^{\prime}$. Consequently, in the Green's function formalism, the reference energy shall be shifted to $i\omega=-d_{0}^{\prime}$ such that the diagonal element of Green's function vanishes and Eq.~(\ref{topo_invariant}) is applicable (see the calculation below). In the noninteracting model described by Eq.~(\ref{BDI_1D_d_vector}), the $d_{0}$ component may be taken to be zero $d_{0}=0$ following the usual convention, and hence $i\omega=0$ in Eq.~(\ref{topo_invariant})\cite{Essin11,Gurarie11}. 

%For the case of electron-phonon interaction, we will demonstrate that $i\omega=0$ will be a suitable choice at which $d_{0}=d_{3}=0$ such that Eq.~(\ref{topo_invariant}) is applicable. 

From Eq.~(\ref{Dyson_equation}) one sees that the full Green's function at $i\omega=-d_{0}^{\prime}$ has the same property and the unperturbed Green's function, namely it only has off-diagonal elements
\begin{eqnarray}
G(k,i\omega=-d_{0}^{\prime})=
\left(
\begin{array}{cc}
0 & \frac{-Q_{k}-\Sigma_{AB}}{|Q_{k}+\Sigma_{AB}|^{2}} \\
\frac{-Q_{k}^{\ast}-\Sigma_{AB}^{\ast}}{|Q_{k}+\Sigma_{AB}|^{2}} & 0
\end{array}
\right)\;,
\label{G_full_E0}
\end{eqnarray}
where $\Sigma_{AB}\equiv\Sigma_{AB}(k,i\omega=-d_{0}^{\prime})=\Sigma_{BA}(k,i\omega=-d_{0}^{\prime})^{\ast}$ is the upper-right off-diagonal element of the $2\times 2$ self-energy matrix at $i\omega=-d_{0}^{\prime}$. Putting Eq.~(\ref{G_full_E0}) into Eq.~(\ref{topo_invariant}), the topological invariant is 
\begin{eqnarray}
{\cal C}&=&\frac{1}{4\pi i}\int_{0}^{2\pi} dk \left\{\frac{-1}{Q_{k}+\Sigma_{AB}}\partial_{k}\left(Q_{k}+\Sigma_{AB}\right)
-h.c.\right\}
\nonumber \\
&=&\frac{1}{2\pi}\int_{0}^{2\pi} dk \;\partial_{k}\varphi\;,
\label{topo_invariant_dphi}
\end{eqnarray} 
where $\varphi_{k}$ is the argument of $Q_{k}+\Sigma_{AB}=|Q_{k}+\Sigma_{AB}|e^{-i\varphi_{k}}$. Thus the topology of the system simply counts the number of times that the phase of $Q_{k}+\Sigma_{AB}$ winds as $k$ changes from $0$ to $2\pi$, which is determined by both the off-diagonal element of the noninteracting Hamiltonian $Q_{k}=d_{1}-id_{2}$ and that of the self-energy $\Sigma_{AB}$. Obviously, the winding number only takes integer values.

%{\cblue (1) Actually I think in 1D the Eqs.~(\ref{topo_invariant}) and (\ref{topo_invariant_dphi}) als works even if the self-energy depends on $E$, i.e., even if $\Sigma_{AB}=\Sigma_{AB}(E)$, in which case one just choose $\Sigma_{AB}(i\omega=\mu)$ to construct the topological invariant. The question is just whether I can still think of $G(i\omega=\mu)=\int\tau\tilde{G}(\tau)$ as the integration of imaginary time Green's function. Should I do SSH model with electron-phonon interaction as an example? }

\begin{figure}[ht]
\begin{center}
\includegraphics[clip=true,width=0.85\columnwidth]{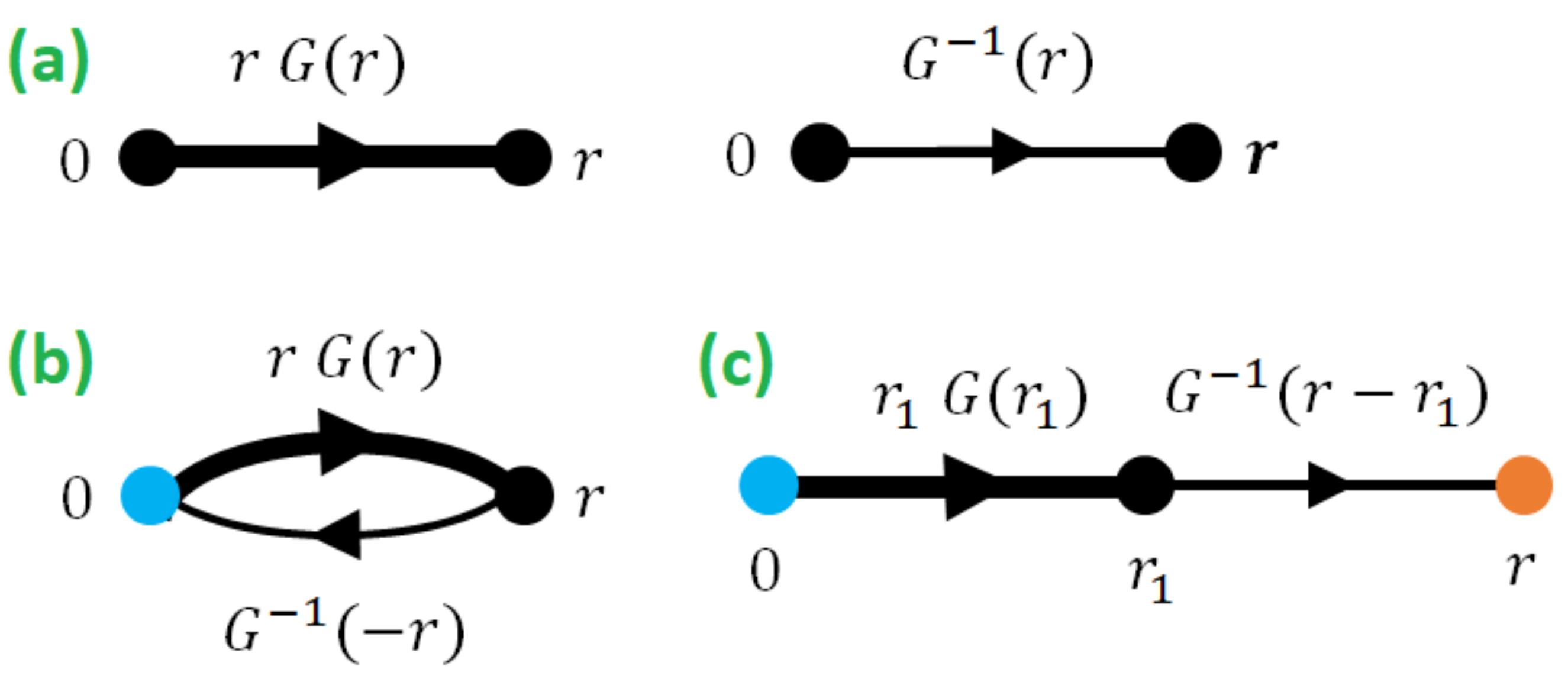}
\caption{ (a) Graphic presentation of the Green's function times the polarization $r\,G(r,i\omega=-d_{0}^{\prime})$ as a thick line and the inverse of the Green's function $G^{-1}(r,i\omega=-d_{0}^{\prime})$ as a thin line. (b) Topological invariant for the interacting 1D class BDI models expressed in terms of the spacetime Green's function. The blue point denotes the origin, and the black points denote the position $r$ that is to be integrated. (c) The correlation function $\lambda_{R}$ presented graphically, which decays with a correlation length that diverges at the topological phase transition. } 
\label{fig:1D_class_BDI_topo_inv_corr_fn_graphics}
\end{center}
\end{figure}

Equation (\ref{topo_invariant_dphi}) indicates that the topological invariant is calculated from the integration of a curvature function, in this case simply the gradient of the phase
\begin{eqnarray}
F(k,{\bf M})\equiv\;\frac{1}{2i}{\rm Tr}\left(\sigma_{3}G^{-1}\partial_{k}G\right)|_{i\omega=-d_{0}^{\prime}}=\partial_{k}\varphi_{k}\;.
\label{curvature_function}
\end{eqnarray}
Assuming that the curvature function displays the divergence discussed in Sec.~\ref{sec:criticality}, which is indeed the case for the 1D class BDI Dirac model with the density-density interaction, a correlation function that characterizes the topological phase transition can be introduced. Consider the Fourier transform between real space and momentum space Green's function
\begin{eqnarray}
G(r,i\omega)&=&\frac{1}{N}\sum_{k}e^{ikr}G(k,i\omega)=\int\frac{dk}{2\pi}e^{ikr}G(k,i\omega)\;,
\nonumber \\
G(k,i\omega)&=&\sum_{r}e^{-ikr}G(r,i\omega)=\int dr\,e^{-ikr}G(r,i\omega)\;,
\nonumber 
\label{Gr_Gk_Fourier_1D}
\end{eqnarray}
and likewisely for $G^{-1}$, where we have set the lattice constant $a=1$ to be unity and dimensionless, and hence the momentum and position are also dimensionless $\left[k\right]=\left[r\right]=1$, and the volume of the system is $v=Na=N$. The topological invariant can then be expressed in terms of the real space Green's function
\begin{eqnarray}
{\cal C}=-\frac{1}{2}\int dr\,{\rm Tr}\left[\sigma_{3}G^{-1}(-r,i\omega)\,r\,G(r,i\omega)\right]_{i\omega=-d_{0}}\;.
\end{eqnarray} 
In addition, the Fourier transform of the curvature function yields a correlation function
\begin{eqnarray}
&&\lambda_{r}=\int_{0}^{2\pi} \frac{dk}{2\pi} \,e^{ikr}\,F(k,{\bf M})
\nonumber \\
&&=-\frac{1}{2}\int dr_{1}\,{\rm Tr}\left[\sigma_{3}G^{-1}(r-r_{1},i\omega)\,r_{1}\,G(r_{1},i\omega)\right]_{i\omega=-d_{0}}\;.
\nonumber \\
\label{lambdaR_1D}
\end{eqnarray}
In this representation, both ${\cal C}$ and $\lambda_{r}$ take the form of a product of the Green's function, its inverse, and the distance it propagates, as depicted
graphically in Fig.~\ref{fig:1D_class_BDI_topo_inv_corr_fn_graphics}. Putting the Ornstein-Zernike form of Eq.~(\ref{Ornstein_Zernike}) into the Fourier transform in Eq.~(\ref{lambdaR_1D}), we see that the correlation function exponentially decays at large distance $\lambda_{r}\propto e^{-r/\xi}$ in either topologically trivial or nontrivial phase, with $\xi$ playing the role of the correlation length. Moreover, the correlation length diverges at the topological phase transition ${\bf M}_{c}$, signifying scale invariance.

\subsection{Su-Schrieffer-Heeger model with nearest-neighbor interaction\label{sec:SSHnn}}

As a concrete example for 1D class BDI, we consider the spinless Su-Schrieffer-Heeger (SSH) model in the presence of a density-density interaction. The noninteracting part of the Hamiltonian is 
\begin{eqnarray}
{\cal H}_{0}&=&\sum_{i}(t+\delta t)c_{Ai}^{\dag}c_{Bi}+(t-\delta t)c_{Ai+1}^{\dag}c_{Bi}+h.c.
\nonumber \\
&=&\sum_{k}Q_{k}c_{Ak}^{\dag}c_{Bk}+Q_{k}^{\ast}c_{Bk}^{\dag}c_{Ak}\;,
\label{SSH_unperturbed}
\end{eqnarray}
where $Q_{k}=(t+\delta t)+(t-\delta t)e^{-ik}$. We examine the nearest-neighbor density-density interaction
\begin{eqnarray}
{\cal H}_{e-e}=V\sum_{i}\left(n_{Ai}n_{Bi}+n_{Bi}n_{Ai+1}\right)\;,
\label{SSH_ee_interaction}
\end{eqnarray}
where $n_{Ii}\equiv c_{Ii}^{\dag}c_{Ii}$, whose Fourier transform takes the form of Eq.~(\ref{general_density_density_interaction}) with $V_{q}=V(1+\cos q)$. The model remains spinless in the presence of the interaction in Eq.~(\ref{SSH_ee_interaction}), in contrast to the spinful version of this model under the influence of Hubbard interactions, which has been investigated previously\cite{Manmana14,Yoshida14}.

Following Appendix \ref{appendix:self_energy}, the one-loop calculation gives the self-energies 
\begin{eqnarray}
&&\Sigma_{AA}(k)=\Sigma_{BB}(k)=V\;,
\nonumber \\
&&\Sigma_{AB}(k)=\frac{1}{2}\sum_{q}V_{q}e^{-i\alpha_{k+q}}
=\left[\Sigma_{BA}(k)\right]^{\ast}\;,
\label{SSHnn_self-energy}
\end{eqnarray}
where the phase $\alpha$ is defined from the diagonal element of the noninteracting Hamiltonian $Q_{k}\equiv|Q_{k}|e^{-i\alpha_{k}}$, and $n_{F}(x)=\theta(-x)$ is the Fermi distribution function that takes the form of a step function at zero temperature. In deriving Eq.~(\ref{SSHnn_self-energy}), we have used Eqs.~(\ref{BDI_1D_d_vector}), (\ref{Hartree_Fock_self_energy}), and the result that follow $G_{0AA}(p,\tau=0)=1/2$ and $G_{0AB}(p+q,\tau=0)=-e^{-i\alpha_{p+q}}/2$. The full Green's function is therefore
\begin{eqnarray}
&&G(k,i\omega)=\frac{1}{(i\omega-V)^{2}+|Q_{k}+\Sigma_{AB}(k)|^{2}}
\nonumber \\
&&\times\left(
\begin{array}{cc}
i\omega-V & Q_{k}+\Sigma_{AB}(k) \\
\left[Q_{k}+\Sigma_{AB}(k)\right]^{\ast} & i\omega-V
\end{array}
\right)\;.
\end{eqnarray}
As explained after Eq.~(\ref{G_full_E0}), at $i\omega=V$ the full Green's function becomes off-diagonal, and the topology simply counts the winding of the phase $\varphi_{k}$ of the off-diagonal element across the BZ
\begin{eqnarray}
&&\varphi(k,{\bf M})=-\arg\left(Q_{k}+\Sigma_{AB}\right)
\nonumber \\
&&=-\arg\left(Q_{k}+\frac{1}{2}\sum_{q}V_{q}e^{-i\alpha_{k+q}}\right)\;.
\label{SSHnn_phik_analytical_result}
\end{eqnarray}
It is then clear that the calculation of topological invariant ${\cal C}$ in Eq.~(\ref{topo_invariant_dphi}) involves two momentum integrations: one integrates over $k$ in Eq.~(\ref{topo_invariant_dphi}) for the topological invariant itself, and the other integrates over $q$ in Eq.~(\ref{SSHnn_phik_analytical_result}) for the one-loop self-energy.

As explain in Sec.~\ref{sec:RG}, we now apply the CRG approach to study the topological phase transitions in the parameter space ${\bf M}=(\delta t,V)$ without explicitly calculating the integration over $k$ in Eq.~(\ref{topo_invariant_dphi}). Since $\partial_{k}\varphi_{k}$ plays the role of the curvature function, the RG equation is that in Eqs.~(\ref{generic_RG_eq}) and (\ref{generic_RG_eq_discrete}) with $F({\bf K},{\bf M})=F(k,{\bf M})=\partial_{k}\varphi_{k}$. Going one step further, since often times the phase $\varphi_{k}=\varphi(k,{\bf M})$ itself is the quantity that is easiest to be solved numerically, one can express the RG equation in terms of $\varphi(k,{\bf M})$ on several mesh points
\begin{eqnarray}
&&\frac{dM_{i}}{dl}=\left(\frac{\Delta M_{i}}{\Delta k^{2}}\right)\left\{\varphi(k_{0}+2\Delta k,{\bf M})-2\varphi(k_{0}+\Delta k,{\bf M})\right.
\nonumber \\
&&\left.+\varphi(k_{0},{\bf M})\right\}
\times\left\{\varphi(k_{0}+\Delta k,{\bf M}+\Delta{\bf M}_{i})\right.
\nonumber \\
&&\left.-\varphi(k_{0},{\bf M}+\Delta{\bf M}_{i})-\varphi(k_{0}+\Delta k,{\bf M})+\varphi(k_{0},{\bf M})\right\}^{-1}\;,
\nonumber \\
\label{1D_generic_RG_eq_phi}
\end{eqnarray}
from which one sees that only five mesh points of $\varphi(k,{\bf M})$ are required to obtain the RG flow at a given ${\bf M}$, which is numerically much more economical compared to a brute force integration of Eq.~(\ref{topo_invariant_dphi}).

\begin{figure}[ht]
\begin{center}
\includegraphics[clip=true,width=0.8\columnwidth]{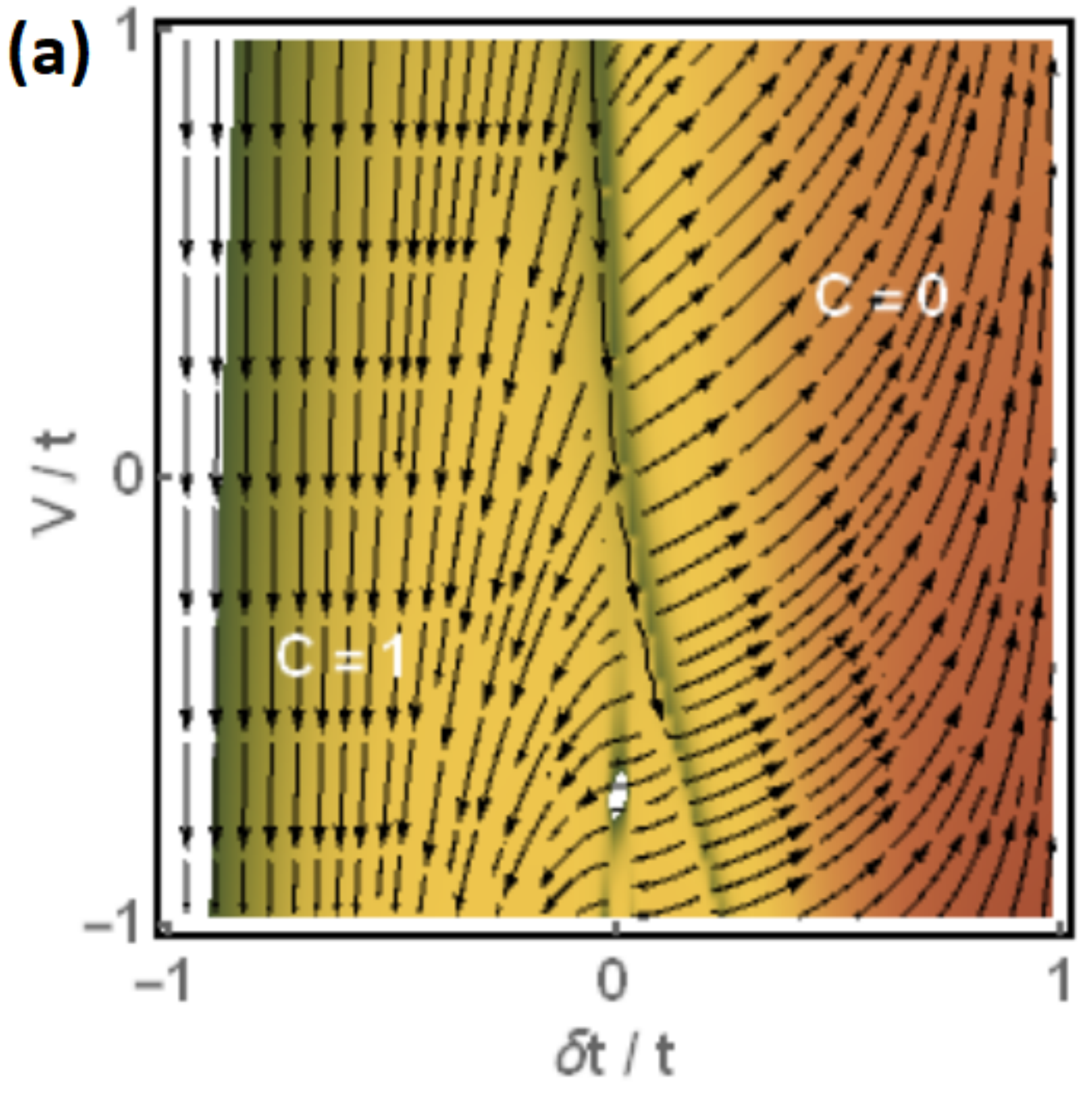}
\includegraphics[clip=true,width=0.99\columnwidth]{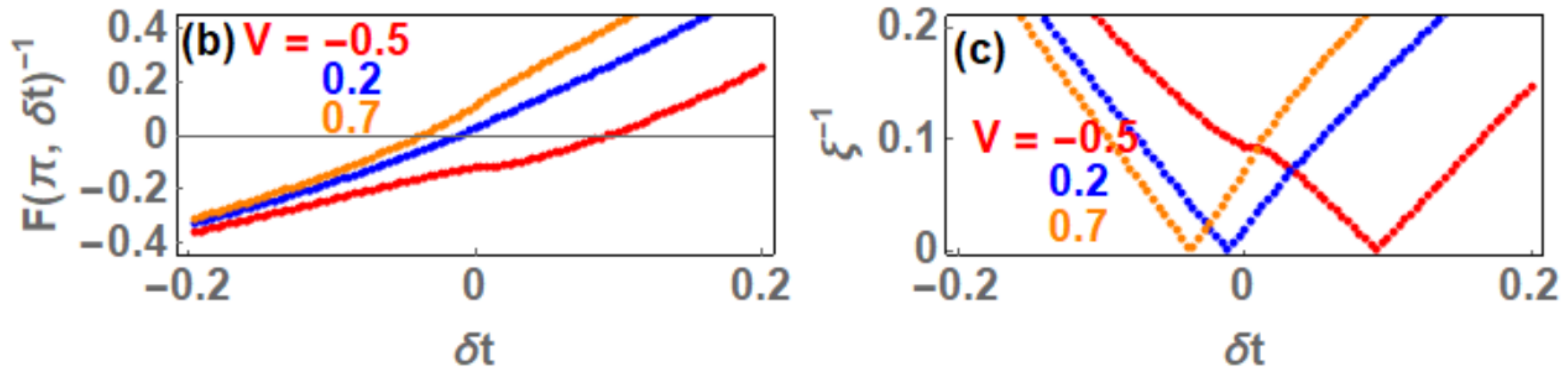}
\includegraphics[clip=true,width=0.99\columnwidth]{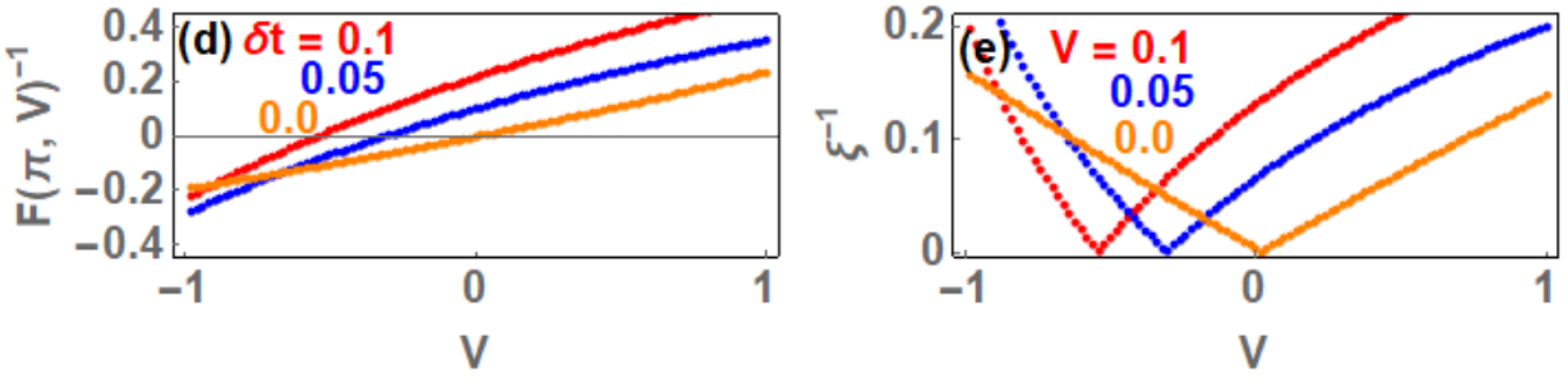}
\caption{ (a) RG flow $d{\bf M}/dl=(d\delta t/dl,dV/dl)$ (black arrows) of SSH model with nearest-neighbor interaction. The $dV/dl$ component is manually reduced by a factor of $0.1$ for the sake of presentation. The log of the flow rate $\ln(|d{\bf M}/dl|)$ is indicated by the color scale, with brown the low flow rate and green the high flow rate. The flow identifies a phase boundary (green line in the middle from which the arrows direct away) between the topologically nontrivial ${\cal C}=1$ and the trivial ${\cal C}=0$ phase. The fixed points seem to locate at ${\bf M}_{f}=(\delta t_{f},V_{f})=(-t,-\infty)$ and $(t,\infty)$ for the two phases, respectively. (b) and (c) show the $F(k_{0}=\pi,{\bf M})^{-1}$ and $\xi^{-1}$ as functions of $\delta t$ at several values of $V$, and (d) and (e) show them as functions of $V$ at several values of $\delta t$. The linear behavior of them as approaching the critical points (where they vanish) indicates the critical exponents in Eq.~(\ref{Fk0_xi_critical_exponent}) are $\gamma\approx\nu\approx 1$.  } 
\label{fig:SSHnn_RGflow}
\end{center}
\end{figure}

The resulting RG flow in the ${\bf M}=(\delta t,V)$ parameter space by using $k_{0}=\pi$ is shown in Fig.~\ref{fig:SSHnn_RGflow}. The RG flow identifies the phase boundary between two topologically distinct phases according to the rule in Eq.~(\ref{identifying_Mc_Mf}). The critical point in the noninteracting limit located at $(\delta t_{c},V)=(0,0)$ separates the topologically trivial ${\cal C}=0$ and nontrivial ${\cal C}=1$ phases, a well-known result for the noninteracting SSH model, and the phase boundary in the interacting model is a continuous line that passes through this critical point. The fixed point for the nontrivial phase seems to locate at ${\bf M}_{f}=(\delta t_{f},V_{f})=(-t,-\infty)$, i.e., with an infinite attractive nearest-neighbor interaction, and for the trivial phase seems to locate at ${\bf M}_{f}=(\delta t_{f},V_{f})=(t,\infty)$, i.e., an infinite repulsive nearest-neighbor interaction. Certainly our Green's function approach is limited within weak coupling $V\apprle t$ so these fixed points are out of the reach of this approach, whose investigation may require other types of numerical approach, such as exact diagonalization.

Figure \ref{fig:SSHnn_RGflow} also shows the critical behavior of $F(k_{0},{\bf M})$ and $\xi$ as approaching the critical point from either $\delta t\rightarrow\delta t_{c}$ or $V\rightarrow V_{c}$ direction. Both $F(k_{0},{\bf M})^{-1}$ and $\xi^{-1}$ clearly demonstrate a linear behavior as approaching the phase boundary from either direction, indicating the critical exponents defined in Eq.~(\ref{Fk0_xi_critical_exponent}) are $\gamma\approx \nu\approx 1$. Besides satisfying the scaling law introduced in Eq.~(\ref{scaling_law}), these exponents coincide with those in the noninteracting limit\cite{Chen17}.

%{\cblue (3) Maybe I'll need Lehmann representation to define correlation function for the weakly interacting cases. }

%{\cblue (4) Green's function RG for topological phase transitions: from $G(\omega=0,k)$ one can construct a topological Hamiltonian $H_{top}$, then one can get its eigenstate, which is just $|\alpha(\omega=0,k)\rangle$, then construct Wannier state by Fourier transform, then correlation function. See Lang's papers. }

\section{2D class A with interactions}

\subsection{Frequency-dependent self-energy\label{sec:Green_function_2D}}

We proceed to consider the 2D spinless fermionic systems in class A that do not preserve any of the three nonspatial symmetries. The ${\bf d}$-vector in the noninteracting model of Eq.~(\ref{general_2by2_Dirac_Hamiltonian}) in this symmetry class satisfies
\begin{eqnarray}
d_{1,2}({\bf k})=-d_{1,2}(-{\bf k})\;,\;\;\;d_{0,3}({\bf k})=d_{0,3}(-{\bf k})\;.
\end{eqnarray}
That is, the noninteracting Dirac Hamiltonian has $d_{1,2}$ component that are odd in ${\bf k}$, and the $d_{0,3}$ component that are even in ${\bf k}$. The $d_{0}$ component is conventionally set to zero, but can be generated by interactions.

 An expression for the topological invariant in terms of full Green's function has been proposed\cite{Niu85,Gurarie11}
\begin{eqnarray}
{\cal C}=&&\frac{\pi}{3}\int_{BZ} \frac{d^{2}{\bf k}}{(2\pi)^{2}}\int_{-\infty}^{\infty}\frac{d\omega}{2\pi}
\nonumber \\
&&\times\epsilon^{abc}{\rm Tr}\left[(G^{-1}\partial_{a}G)(G^{-1}\partial_{b}G)(G^{-1}\partial_{c}G)\right]
\label{2D_C_interacting}
\end{eqnarray}
where $\epsilon^{abc}$ is antisymmetric in the three indices $\left\{a,b,c\right\}=\left\{\omega,k_{x},k_{y}\right\}$, and $G\equiv G({\bf k},i\omega)$ is the interaction-dressed single-particle Green's function. We now examine the integrand in this expression using the renormalized ${\bf d}^{\prime}$-vector in Eqs.~(\ref{2D_G_dprime}) and (\ref{renormalized_d_vector_mu}). Keeping in mind that $d_{i}^{\prime}$ is a function of both momentum and frequency, and hence $\partial_{\omega}d_{i}^{\prime}$ and $\partial_{k_{j}}d_{i}^{\prime}$ are both nonzero, a straight forward expansion of the integrand in Eq.~(\ref{2D_C_interacting}) gives
\begin{eqnarray}
&&\frac{\pi}{3}\epsilon^{abc}{\rm Tr}\left[(G^{-1}\partial_{a}G)(G^{-1}\partial_{b}G)(G^{-1}\partial_{c}G)\right]_{\left\{a,b,c\right\}=\left\{\omega,k_{x},k_{y}\right\}}
\nonumber \\
&&=\frac{4\pi i}{\left[(i\omega+d_{0}^{\prime})^{2}-d^{\prime 2}\right]^{2}}\left\{
-i\epsilon^{abc}d_{a}^{\prime}\partial_{x}d_{b}^{\prime}\partial_{y}d_{c}^{\prime}
|_{\left\{a,b,c\right\}=\left\{1,2,3\right\}}
\right.
\nonumber \\
&&+\epsilon^{abcd}d_{a}^{\prime}\partial_{\omega}d_{b}^{\prime}\partial_{x}d_{c}^{\prime}\partial_{y}d_{d}^{\prime}
|_{\left\{a,b,c,d\right\}=\left\{0,1,2,3\right\}}
\nonumber \\
&&\left.+i\omega\epsilon^{abc}\partial_{\omega}d_{a}^{\prime}\partial_{x}d_{b}^{\prime}\partial_{y}d_{c}^{\prime}
|_{\left\{a,b,c\right\}=\left\{1,2,3\right\}}\right\},
\label{TrGGGGGG_to_dprime_to_FkM}
\nonumber \\
&&=F({\bf K},{\bf M})\;,
\end{eqnarray}
where $\epsilon^{abc}$ and $\epsilon^{abcd}$ are antisymmetric in permutations of two neighboring indices. In the spirit of Eq.~(\ref{topo_invariant}), this integrand is treated as a curvature function $F({\bf K},{\bf M})$ defined in the momentum-frequency space ${\bf K}=(\omega,k_{x},k_{y})$, and for the models examined in this article, we find that it exhibits the critical behavior described in Sec.~\ref{sec:criticality}.

To demonstrate the critical behavior, we first examine the following continuous limit of the noninteracting lattice model $d_{i}^{\prime}\rightarrow d_{i}$\cite{Bernevig13}
\begin{eqnarray}
&&d_{0}=0\;,\;\;\;d_{1}=k_{x},\;\;\;d_{2}=k_{y},\;\;\;d_{3}=M,
\nonumber \\
&&\epsilon^{abc}d_{a}\partial_{x}d_{b}\partial_{y}d_{c}=M.
\end{eqnarray}
In this limit, the second and third term in Eq.~(\ref{TrGGGGGG_to_dprime_to_FkM}) drops out because $\partial_{\omega}d_{i}=0$, and hence
\begin{eqnarray}
&&\frac{\pi}{3}\epsilon^{abc}{\rm Tr}\left[(G_{0}^{-1}\partial_{a}G_{0})(G_{0}^{-1}\partial_{b}G_{0})(G_{0}^{-1}\partial_{c}G_{0})\right]
\nonumber \\
&&=\frac{4\pi M}{(\omega^{2}+k^{2}+M^{2})^{2}}
\approx\frac{\tilde{F}({\bf K}_{0},M)}{1+\xi^{2}K^{2}}\;,
\label{FkM_2D_class_A_noninteracting_limit}
\end{eqnarray}
where in the last line we have expanded with $K\ll 1$ and used $d^{2}=k^{2}+M^{2}$. The result suggests that the curvature function takes the Ornstein-Zernike form around the HSP ${\bf K}_{0}=(0,0,0)$ in all three directions in the momentum-frequency space, with the height of the Lorentzian $\tilde{F}({\bf K}_{0},M)=4\pi/M^{3}$ that diverges with critical exponent $\gamma=3$ and changes sign at the critical point $M_{c}=0$, and the inverse of the Lorentzian width in all three directions $\xi_{i}=\sqrt{2}/|M|$ diverges with critical exponent $\nu_{i}=1$. Thus the noninteracting Dirac model treated within the Green's function formalism satisfies the critical behavior and the scaling law $\gamma=\nu_{t}+\nu_{x}+\nu_{y}$ introduced in Sec.~\ref{sec:criticality}.

\begin{figure}[ht]
\begin{center}
\includegraphics[clip=true,width=0.8\columnwidth]{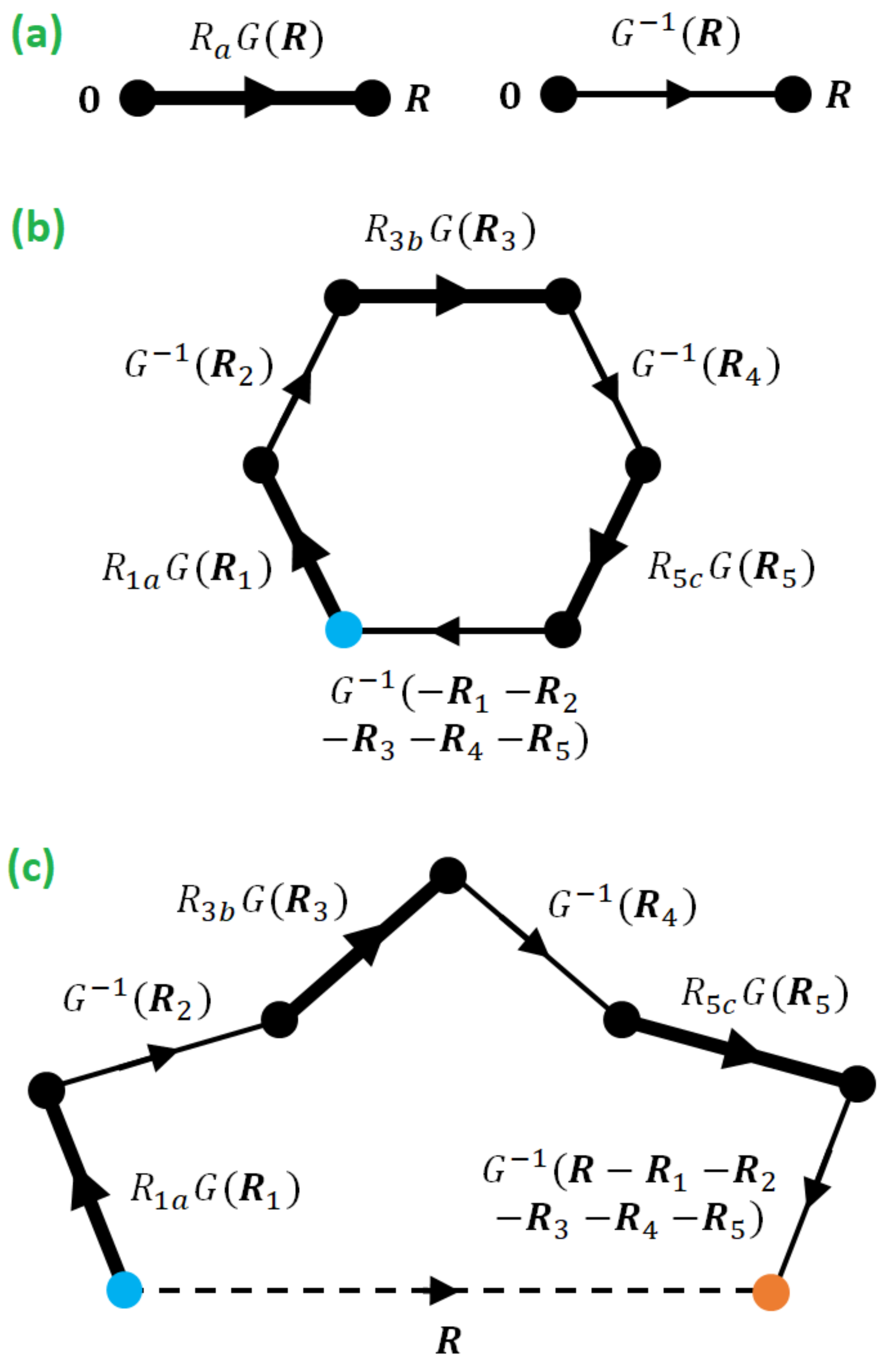}
\caption{ (a) Graphic presentation of the spacetime Green's function times a distance $R_{a}G({\bf R})$ as a thick line and the inverse of the spacetime Green's function $G^{-1}({\bf R})$ as a thin line. (b) Topological invariant for the interacting 2D class A models expressed in terms of the spacetime Green's function. The blue point denotes the origin, and the 5 black points denote the positions ${\bf R}_{1}\sim{\bf R}_{5}$ that are to be integrated. (c) The correlation function $\lambda_{\bf R}$ presented graphically, which decays with a correlation length in all the three spacetime directions. } 
\label{fig:2D_class_A_topo_inv_corr_fn_graphics}
\end{center}
\end{figure}

Consider the Fourier transform of the Green's function
\begin{eqnarray}
G({\bf R})&=&\int\frac{d^{3}{\bf K}}{(2\pi)^{3}}\,e^{i{\bf K\cdot R}}\,G({\bf K})\;,
\nonumber \\
G({\bf K})&=&\int d^{3}{\bf R}\,e^{-i{\bf K\cdot R}}\,G({\bf R})\;,
\end{eqnarray}
and likewisely for $G^{-1}$, where $e^{i{\bf K}\cdot{\bf R}}=e^{ik_{x}r_{x}+ik_{y}r_{y}-i\omega t}$, $\int d^{3}{\bf K}\equiv\int_{BZ} d^{2}{\bf k}\int_{-\infty}^{\infty} d\omega$, and ${\bf R}=(-t,R_{x},R_{y})$ is the spacetime coordinate. A direct Fourier transform of  Eq.~(\ref{2D_C_interacting}) gives the topological invariant expressed in terms of spacetime Green's functions
\begin{eqnarray}
{\cal C}&=&\frac{i\pi}{3}\left(\prod_{i=1}^{5}\int d^{3}{\bf R}_{i}\right)\epsilon^{abc}
\nonumber \\
&&\times{\rm Tr}\left[\right.G^{-1}(-{\bf R}_{1}-{\bf R}_{2}-{\bf R}_{3}-{\bf R}_{4}-{\bf R}_{5})R_{1a}G({\bf R}_{1})
\nonumber \\
&&\times\left.G^{-1}({\bf R}_{2})R_{3b}G({\bf R}_{3})
G^{-1}({\bf R}_{4})R_{5c}G({\bf R}_{5})\right].
\end{eqnarray}
Following Eq.~(\ref{general_Fourier_trans_lambdaR}), the correlation function is introduced from the Fourier transform of the curvature function
\begin{eqnarray}
&&\lambda_{\bf R}=\int_{BZ} \frac{d^{2}{\bf k}}{(2\pi)^{2}}\int_{-\infty}^{\infty}\frac{d\omega}{2\pi}\,e^{i{\bf K}\cdot{\bf R}}F({\bf K},{\bf M})
\nonumber \\
&&=\frac{i\pi}{3}\left(\prod_{i=1}^{5}\int d^{3}{\bf R}_{i}\right)\epsilon^{abc}
\nonumber \\
&&\times{\rm Tr}\left[\right.G^{-1}({\bf R}-{\bf R}_{1}-{\bf R}_{2}-{\bf R}_{3}-{\bf R}_{4}-{\bf R}_{5})R_{1a}G({\bf R}_{1})
\nonumber \\
&&\times\left.G^{-1}({\bf R}_{2})R_{3b}G({\bf R}_{3})
G^{-1}({\bf R}_{4})R_{5c}G({\bf R}_{5})\right]\;.
\end{eqnarray}
Similar to those for 1D class BDI models discussed in Sec.~\ref{sec:1D_class_BDI}, the topological invariant and the correlation function in this representation are a product of the spacetime Green's functions propagated over several segments, as graphically presented in Fig.~\ref{fig:2D_class_A_topo_inv_corr_fn_graphics}.

%Since the 2D model breaks TR symmetry, the topological invariant of the noninteracting model is the Hall conductance calculated from the integration of Berry curvature, which equivalently counts the skyrmion number of the ${\bf d}=(d_{1},d_{2},d_{3})$ vector. We shall briefly review the Berry curvature and the skyrmion number, from which the physical picture of the topological invariant expressed in terms of Green's function will become more transparent.

In the presence of weak interactions, we use Dyson's equation, Eq.~(\ref{Dyson_equation}), to obtain the full Green's function up to one-loop. The $\left(G_{0}\Sigma G\right)_{IJ}$ part of the Dyson's equation is detailed in Appendix \ref{appendix:self_energy}. The calculation of self-energy now requires a 2D integral over momentum. Together with the 3D integral in Eq.~(\ref{2D_C_interacting}), evidently the calculation of topological invariant ${\cal C}$ at a given tuning parameter ${\bf M}$ requires a total of 5D integration, obviously a tedius task. We will demonstrate how CRG circumvents the 3D integral in Eq.~(\ref{2D_C_interacting}) by using a concrete example in the next section.

%We now demonstrate how the RG approach proposed in Sec.~\ref{sec:RG} circumvents this problem. The RG flow in Eqs.~(\ref{generic_RG_eq}) and (\ref{generic_RG_eq_discrete}) requires to calculate the curvature function at 3 mesh points $F({\bf K}_{0},{\bf M})$, $F({\bf K}_{0}+\Delta{\bf K},{\bf M})$, and $F({\bf K}_{0},{\bf M}+\Delta{\bf M}_{i})$, where the small displacement $\Delta{\bf K}$ away from the HSP ${\bf K}_{0}$ can be along any direction in the momentum-frequency space in which the correlation length or correlation time $\xi_{i}$ vanishes as ${\bf M}\rightarrow{\bf M}_{c}$. The RG approach therefore is a numerically very powerful technique, since it requires to calculate only few points without a brute force integration over the entire momentum-frequency space as demanded by Eq.~(\ref{2D_C_interacting}).

\begin{figure}[ht]
\begin{center}
\includegraphics[clip=true,width=0.7\columnwidth]{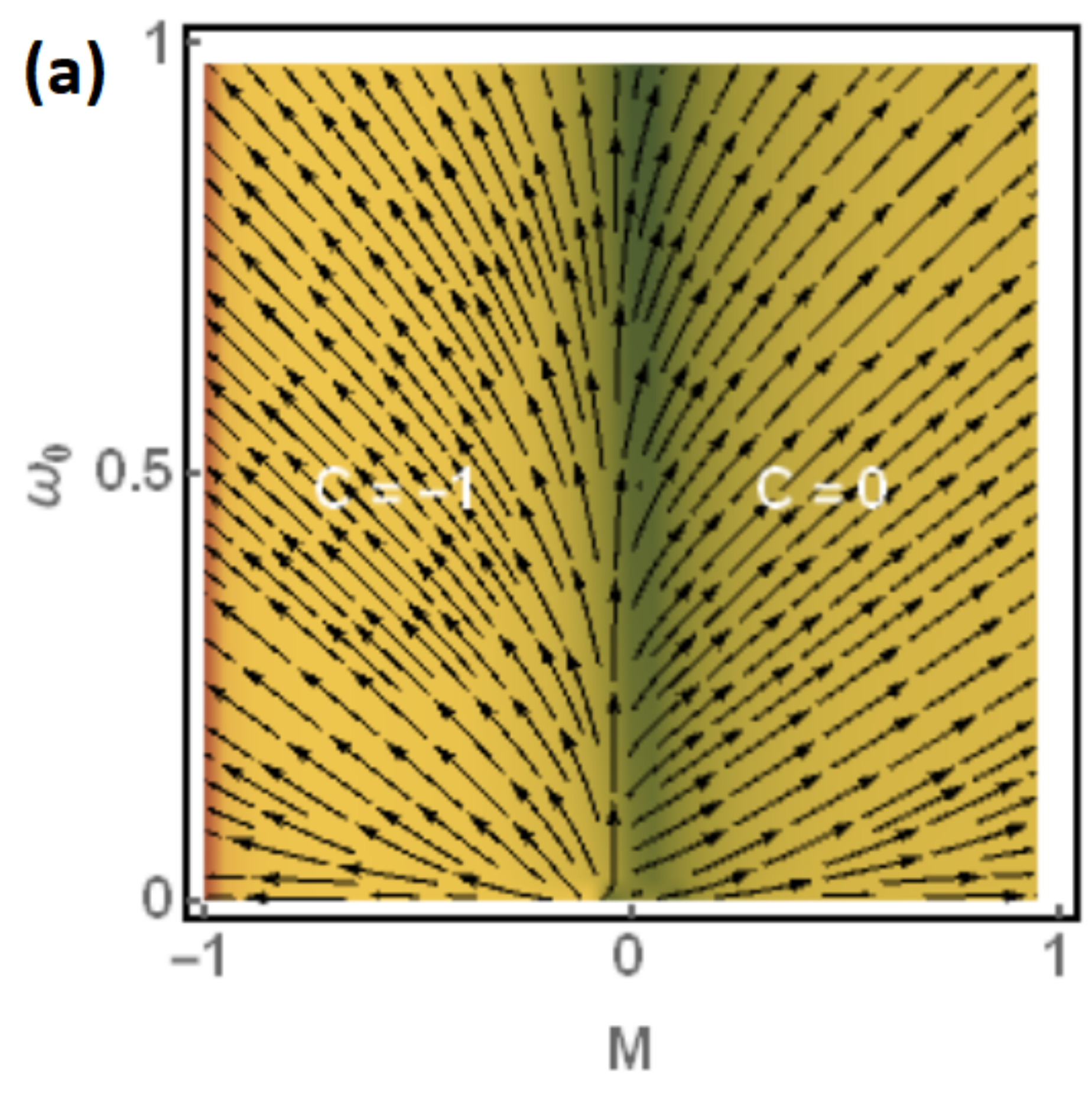}
\includegraphics[clip=true,width=0.99\columnwidth]{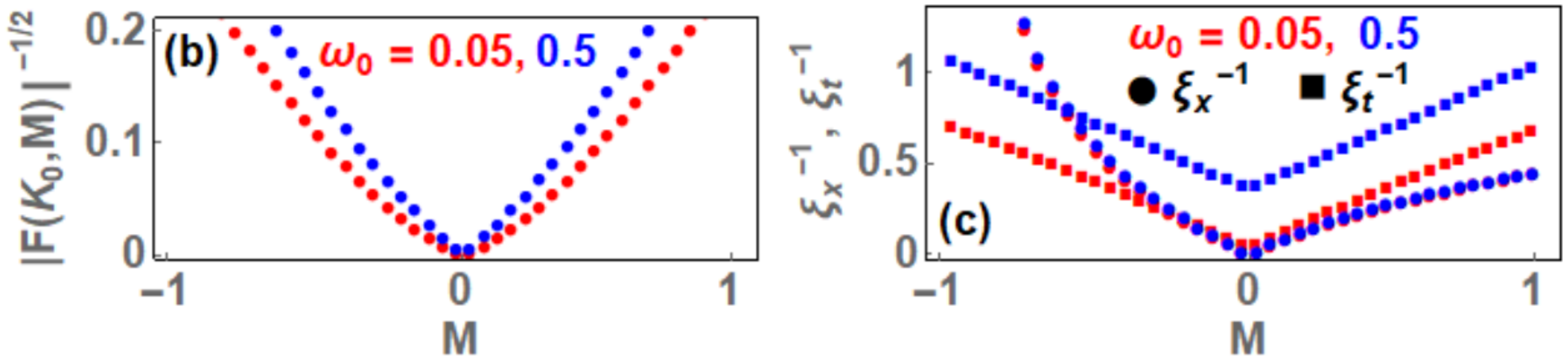}
\caption{ (a) The RG flow $d{\bf M}/dl$ (black arrows) of 2D Chern insulator with electron-phonon interaction in the ${\bf M}=(M,\omega_{0})$ parameter space. The flow component $d\omega_{0}/dl$ is reduced by a factor of $0.1$ for the sake of presentation. The color scale indicates the high (green) and low (brown) flow rate on log scale $\ln|d{\bf M}/dl|$. The flow identifies a phase boundary (green line) between the topologically nontrivial ${\cal C}=-1$ and trivial ${\cal C}=0$ phase that is independent from the phonon frequency $\omega_{0}$. (b) and (c) show the $|F({\bf K}_{0},M)|^{1/2}$ and $\left\{\xi_{x}^{-1},\xi_{t}^{-1}\right\}$. At large $\omega_{0}$, $|F({\bf K}_{0},M)|^{1/2}$ shows linear behavior near the critical point $M_{0}=0$, indicating the critical exponent $\gamma\approx 2$, where as at small $\omega_{0}$ it deviates from linear and the exponent approaches $\gamma\approx 3$. The linear behavior of $\left\{\xi_{x}^{-1},\xi_{t}^{-1}\right\}$ near the critical point indicates the critical exponent $\nu_{x}\approx\nu_{t}\approx 1$. In addition, $\xi_{t}^{-1}$ features an offset at the critical point that is linear in $\omega_{0}$, as described by Eq.~(\ref{xi_ephonon_critical_behavior}). } 
\label{fig:Chern_ephonon_RGflow}
\end{center}
\end{figure}

\subsection{2D Chern insulator with electron-phonon interaction}

As an example of frequency-dependent self-energy, we study 2D Chern insulator in the presence of electron phonon interaction ${\cal H}={\cal H}_{0}+{\cal H}_{e-ph}$. The unperturbed lattice Hamiltonian of 2D class A in momentum space is described by Eq.~(\ref{general_2by2_Dirac_Hamiltonian}) with\cite{Bernevig13} 
\begin{eqnarray}
&&d_{0}=0,\;d_{1}=\sin k_{x},\;d_{2}=\sin k_{y},
\nonumber \\
&&d_{3}=M+2-\cos k_{x}-\cos k_{y}\;,
\label{2D_Chern_H0}
\end{eqnarray}
and the electron-phonon interaction is that in Eq.~(\ref{general_electron_phonon_interaction}). After calculating the self-energy using the formula in Appendix \ref{appendix:self_energy}, we calculate the renormalized ${\bf d}^{\prime}$-vector and its derivatives according to Eq.~(\ref{renormalized_d_vector_mu}), and subsequently put them into Eq.~(\ref{TrGGGGGG_to_dprime_to_FkM}) to obtain the curvature function. Numerically, we varify that the curvature function indeed takes the Ornstein-Zernike form of Eq.~(\ref{Ornstein_Zernike}) is all three directions in the momentum-frequency space. We will treat the mass term in the Dirac Hamiltonian and the optical phonon frequency as tuning parameters ${\bf M}=(M,\omega_{0})$.

To study the topological phase transition in the ${\bf M}=(M,\omega_{0})$ parameter space, we apply the CRG technique in Sec.~\ref{sec:RG} by putting the curvature function in Eq.~(\ref{TrGGGGGG_to_dprime_to_FkM}) into Eq.~(\ref{RG_procedure_general}), without explicitly performing the 3D integral of Eq.~(\ref{2D_C_interacting}). The resulting RG flow is shown in Fig.~\ref{fig:Chern_ephonon_RGflow}, which features a phase boundary between the topological nontrivial ${\cal C}=-1$ and trivial ${\cal C}=0$ phase that passes through the known critical point at the noninteracting limit ${\bf M}_{c}=(0,0)$. Interestingly, at a general phonon frequency $\omega_{0}$ the critical point is ${\bf M}_{c}=(0,\omega_{0})$ at any polar coupling strength $u$ in Eq.~(\ref{general_electron_phonon_interaction}), meaning that the topological phase transition is not influenced by the phonon frequency, at least in the weak coupling regime.

However, the phonon frequency influences the critical exponents. We found that the critical behavior of the correlation length $\left\{\xi_{x},\xi_{y}\right\}$ and correlation time $\xi_{t}$ is described by, having in mind that $M_{c}=0$, 
\begin{eqnarray}
\xi_{x}\approx\frac{1}{c_{x}|M|}\;,
\;\;\xi_{y}\approx\frac{1}{c_{y}|M|}\;,
\;\;\xi_{t}\approx\frac{1}{c_{t\omega}\omega_{0}+c_{tM}|M|}\;,\;
\label{xi_ephonon_critical_behavior}
\end{eqnarray}
where the $\left\{c_{x},c_{y},c_{t\omega},c_{tM}\right\}$ are phenomenological fitting parameters. The correlation lengths along the two momentum directions diverge with critical exponents $\nu_{x}=\nu_{y}=1$. The correlation time, on the other hand, displays a more complicated dependence. Equation (\ref{xi_ephonon_critical_behavior}) indicates that in the noninteracting limit $\omega_{0}=0$, the correlation time $\xi_{t}$ has critical exponent $\nu_{t}=1$, consistent with that described after Eq.~(\ref{FkM_2D_class_A_noninteracting_limit}). However, at finite electron-phonon interaction, $\xi_{t}$ no longer diverges at $M\rightarrow M_{c}=0$ but saturates to a constant that is inversely proportional to the phonon frequency $\omega_{0}$, and consequently assigned a critical exponent $\nu_{t}=0$. Therefore according to the scaling law $\gamma=\nu_{x}+\nu_{y}+\nu_{t}$, the critical exponent for the extremum of curvature function should be $\gamma=2$ for the interacting case, and $\gamma=3$ for the noninteracting limit, which are indeed varified numerically and presented in Fig.~\ref{fig:Chern_ephonon_RGflow}. We are lead to conclude that in the presence of electron-phonon interaction, the critical exponents $\left\{\gamma,\nu_{t}\right\}$ are changed, and as $M\rightarrow M_{c}$ the scale invariance does not manifest along the time-direction but only along the two spatial directions. Note that because the curvature function does not diverge in the frequency direction, $\delta{\bf K}$ has to be chosen along either $\delta k_{x}$ or $\delta k_{y}$ in the CRG procedure.

%{\cblue (2) The above analysis implies that although the phonon does not influence the topology, it changes the critical exponents. But this is strange, this means we have like a first-order phase transition at the frequency direction, that the curvature function as a function of frequency changes discontinuously across $M_{c}$. Need to check this carefully.  }

%{\cblue (3) Strange, can this finite $\xi_{t}$ at $M\rightarrow M_{c}$ mean that the transition at $M_{c}$ becomes first order? Because the curvature function does not shrink into an infinitely narrow peak in the frequency direction but just abruptly flips sign across $M_{c}$, indicating a first-order transition? Check this. If this is true then say it's probably not a big deal because in a superconductor, the transition at $T_{c}$ is second-order at zero magnetic field, but at finite magnetic field it becomes first order, as in Manfred's lecture note. Mention this. }

%{\cblue (1) Plot the spectral function at a given electron-phonon interaction, could be instructive to show that gap is not a well defined concept. }

\subsection{Frequency-independent self-energy}

We proceed to address the situation when self-energy is frequency-independent, since it is relevant to several types of density-density interaction. In this case, the ${\bf d}^{\prime}$-vector in Eq.~(\ref{renormalized_d_vector_mu}) is frequency-independent and hence $\partial_{\omega}d_{i}^{\prime}=0$, following which the form of the curvature function in Eq.~(\ref{TrGGGGGG_to_dprime_to_FkM}) is greatly simplified 
\begin{eqnarray}
&&\frac{\pi}{3}\epsilon^{abc}{\rm Tr}\left[(G_{0}^{-1}\partial_{a}G_{0})(G_{0}^{-1}\partial_{b}G_{0})(G_{0}^{-1}\partial_{c}G_{0})\right]
\nonumber \\
&&=\frac{4\pi}{\left[-(i\omega+d_{0}^{\prime})^{2}+d^{\prime 2}\right]^{2}}\epsilon^{abc}d_{a}^{\prime}\partial_{x}d_{b}^{\prime}\partial_{y}d_{c}^{\prime}\;.
\label{TrGdGGdGGdG_to_Fxy}
\end{eqnarray}
As a result, the form of correlation function and the topological invariant expressed in terms of real space Green's function is simplified. The integration over frequency in Eq.~(\ref{2D_C_interacting}) can be performed analytically
\begin{eqnarray}
\int_{-\infty}^{\infty}\frac{d\omega}{\left[-(i\omega+d_{0}^{\prime})^{2}+d^{\prime 2}\right]^{2}}=\frac{\pi}{2d^{\prime 3}}\;\;({\rm assuming}\;|d_{0}^{\prime}|<d^{\prime}).
\nonumber \\
\end{eqnarray}
The assumption of $|d_{0}^{\prime}|<d^{\prime}$ is reasonable, since $d_{0}^{\prime}$ comes from the self-energy according to Eq.~(\ref{renormalized_d_vector_mu}), and in the weak coupling regime it should be smaller than the renormalized band width $d^{\prime}$. The topological invariant in Eq.~(\ref{2D_C_interacting}) then takes the form
\begin{eqnarray}
&&{\cal C}=\frac{1}{4\pi}\int_{BZ}d^{2}{\bf k}\frac{\epsilon^{abc}}{d^{\prime 3}}d_{a}^{\prime}\partial_{x}d_{b}^{\prime}\partial_{y}d_{c}^{\prime}|_{\left\{a,b,c\right\}=\left\{1,2,3\right\}}
\nonumber \\
&&=\frac{1}{4\pi}\int_{BZ} d^{2}{\bf k}\,{\hat{\bf d}}^{\prime}\cdot\left(\partial_{x}{\hat{\bf d}}^{\prime}\times\partial_{y}{\hat{\bf d}}^{\prime}\right)|_{\left\{a,b,c\right\}=\left\{1,2,3\right\}}\;.\;\;\;\;\;
\label{2D_C_Fxy_dprime}
\end{eqnarray}
The physical meaning of the topological invariant becomes clear: it simply counts the skyrmion number of the self-energy-renormalized $(d_{1}^{\prime},d_{2}^{\prime},d_{3}^{\prime})$-vector. Provided $|d_{0}^{\prime}|<d^{\prime}$, the $d_{0}^{\prime}$ component does not influece the topology. In the noninteracting limit $(d_{1}^{\prime},d_{2}^{\prime},d_{3}^{\prime})\rightarrow(d_{1},d_{2},d_{3})$, this skyrmion number is known to coincide with the Hall conductance\cite{Bernevig13}. Interaction modifies the profile of $(d_{1}^{\prime},d_{2}^{\prime},d_{3}^{\prime})$ in momentum space, yet the skyrmion number always takes an integer value.

\begin{figure}[ht]
\begin{center}
\includegraphics[clip=true,width=0.8\columnwidth]{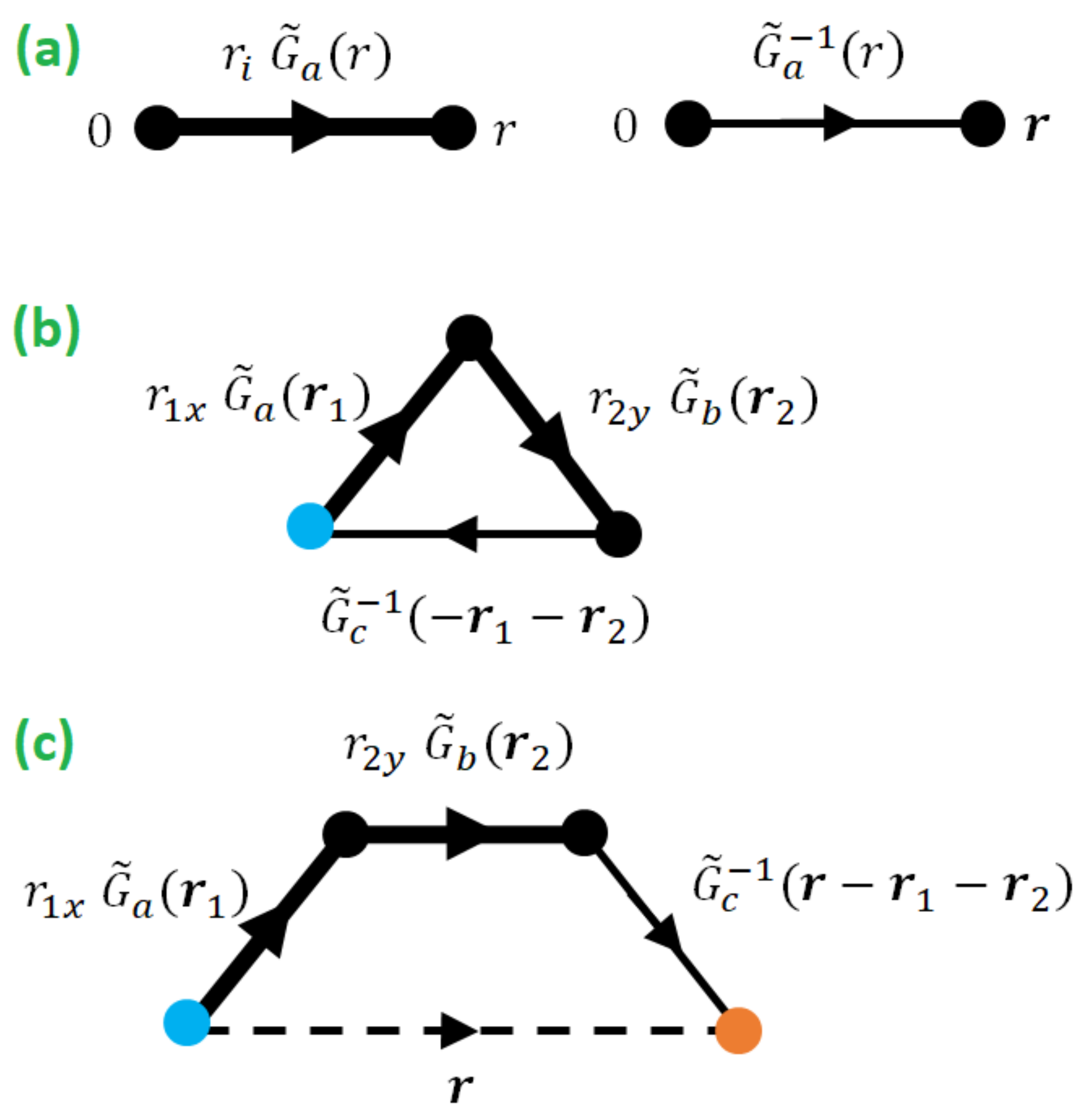}
\caption{ Graphic presentation of several quantities for 2D class A models with frequency-independent self-energy. (a) The frequency-integrated real space Green's function times a distance $r_{i}G({\bf r})$ as a thick line and the inverse of the frequency-integrated Green's function $G^{-1}({\bf r})$ as a thin line. (b) Topological invariant expressed in terms of the real space Green's function. The blue point denotes the origin, and the 2 black points denote the positions that are to be integrated. (c) The correlation function $\lambda_{\bf r}$ presented graphically, which decays with a correlation length. } 
\label{fig:2D_class_A_freq_ind_topo_inv_corr_fn_graphics}
\end{center}
\end{figure}

For this case of frequency-independent self-energy, we discuss the criticality near topological phase transitions in the following manner according to the argument in Sec.~\ref{sec:criticality}. We observe that if we integrate out the frequency variable of the Green's function in Eq.~(\ref{2D_G_dprime}), the result only depends on the ${\bf d}^{\prime}$-vector
\begin{eqnarray}
G({\bf k})\equiv\int\frac{d\omega}{2\pi}G({\bf k},i\omega)
=-\frac{1}{2}\left(
\begin{array}{cc}
-1+\frac{d_{3}^{\prime}}{d^{\prime}} & \frac{d_{1}^{\prime}-id_{2}^{\prime}}{d^{\prime}} \\
\frac{d_{1}^{\prime}+id_{2}^{\prime}}{d^{\prime}} &
-1-\frac{d_{3}^{\prime}}{d^{\prime}}
\end{array}
\right),
\nonumber \\
\label{Gkw_integrate_out_w}
\end{eqnarray}
after substituting Eq.~(\ref{2D_G_dprime}). Consequently, the ${\bf d}^{\prime}$-vector can be written in terms of $G({\bf k})$
\begin{eqnarray}
-d_{1}^{\prime}/d^{\prime}&=&G_{AB}({\bf k})+G_{BA}({\bf k})\equiv \tilde{G}_{1}({\bf k})\;
\nonumber \\
id_{2}^{\prime}/d^{\prime}&=&G_{AB}({\bf k})-G_{BA}({\bf k})\equiv \tilde{G}_{2}({\bf k})\;
\nonumber \\
-d_{3}^{\prime}/d^{\prime}&=&G_{AA}({\bf k})-G_{BB}({\bf k})\equiv \tilde{G}_{3}({\bf k})\;.
\label{dprime_Gi}
\end{eqnarray}
In terms of this newly defined, frequency-independent Green's function elements $\left\{\tilde{G}_{1},\tilde{G}_{2},\tilde{G}_{3}\right\}$, the topological invariant in Eq.~(\ref{2D_C_Fxy_dprime}) is
\begin{eqnarray}
&&{\cal C}=\int\frac{d^{2}{\bf k}}{(2\pi)^{2}}F({\bf k},{\bf M})\;,
\nonumber \\
&&F({\bf k},{\bf M})=\frac{\pi}{i}\epsilon^{abc}\tilde{G}_{a}({\bf k})\partial_{x}\tilde{G}_{b}({\bf k})\partial_{y}\tilde{G}_{c}({\bf k})\;,
\label{2D_C_freq_ind_GaGbGc}
\end{eqnarray}
where the frequency-independent curvature function $F({\bf k},{\bf M})$ is introduced in terms of $\tilde{G}_{a}({\bf k})$, whose profile in momentum space obviously depends on the interacting or noninteracting parameters ${\bf M}=(M_{1},M_{2}...)$ in the Hamiltonian. We then consider the Fourier transform 
\begin{eqnarray}
\tilde{G}({\bf k})=\sum_{\bf r}e^{-i{\bf k\cdot r}}\tilde{G}({\bf r})=\int d^{2}{\bf r}\,e^{-i{\bf k\cdot r}}\tilde{G}({\bf r})\;,
\nonumber \\
\tilde{G}({\bf r})=\frac{1}{N}\sum_{\bf k}e^{i{\bf k\cdot r}}\tilde{G}({\bf k})=\int \frac{d^{2}{\bf k}}{\left(2\pi\right)^{2}}e^{i{\bf k\cdot r}}\tilde{G}({\bf k})\;.
\label{Gk_Gr_Fourier}
\end{eqnarray}
to define the real space counterpart of $\tilde{G}({\bf k})$. In terms of these real space functions, the topological invariant in Eq.~(\ref{2D_C_freq_ind_GaGbGc}) reads, after a Fourier transform,
\begin{eqnarray}
{\cal C}&=&i\pi\int d^{2}{\bf r}_{1}\int d^{2}{\bf r}_{2}
\nonumber \\
&&\times\epsilon^{abc}\tilde{G}_{a}(-{\bf r}_{1}-{\bf r}_{2})r_{1x}\tilde{G}_{b}({\bf r}_{1})r_{2y}\tilde{G}_{c}({\bf r}_{2})\;.
\end{eqnarray}
We proceed to follow the principle mentioned in Sec.~\ref{sec:criticality} to introduce the correlation function as the Fourier transform of the curvature function. In this case the result is
\begin{eqnarray}
&&\lambda_{\bf r}=\int\frac{d^{2}{\bf k}}{(2\pi)^{2}}e^{i{\bf k\cdot r}}F({\bf k},{\bf M})
\nonumber \\
&&=i\pi\int d^{2}{\bf r}_{1}\int d^{2}{\bf r}_{2}
\nonumber \\
&&\times\epsilon^{abc}\tilde{G}_{a}({\bf r}-{\bf r}_{1}-{\bf r}_{2})r_{1x}\tilde{G}_{b}({\bf r}_{1})r_{2y}\tilde{G}_{c}({\bf r}_{2})\;,
\nonumber \\
\label{lambdaR_2D}
\end{eqnarray}
as present graphically in Fig.~\ref{fig:2D_class_A_freq_ind_topo_inv_corr_fn_graphics}. Note that in these expressions, $\tilde{G}_{i}({\bf r})$ is a function calculated from the matrix elements of $G({\bf k},i\omega)$ integrated out the frequency variable and then Fourier transformed to real space, according to Eqs.~(\ref{Gkw_integrate_out_w}), (\ref{dprime_Gi}), and (\ref{Gk_Gr_Fourier}), and ${\cal C}$ and $\lambda_{\bf r}$ are products of $\tilde{G}_{i}({\bf r})$ and its inverse propagated over several segments.

\begin{figure}[ht]
\begin{center}
\includegraphics[clip=true,width=0.7\columnwidth]{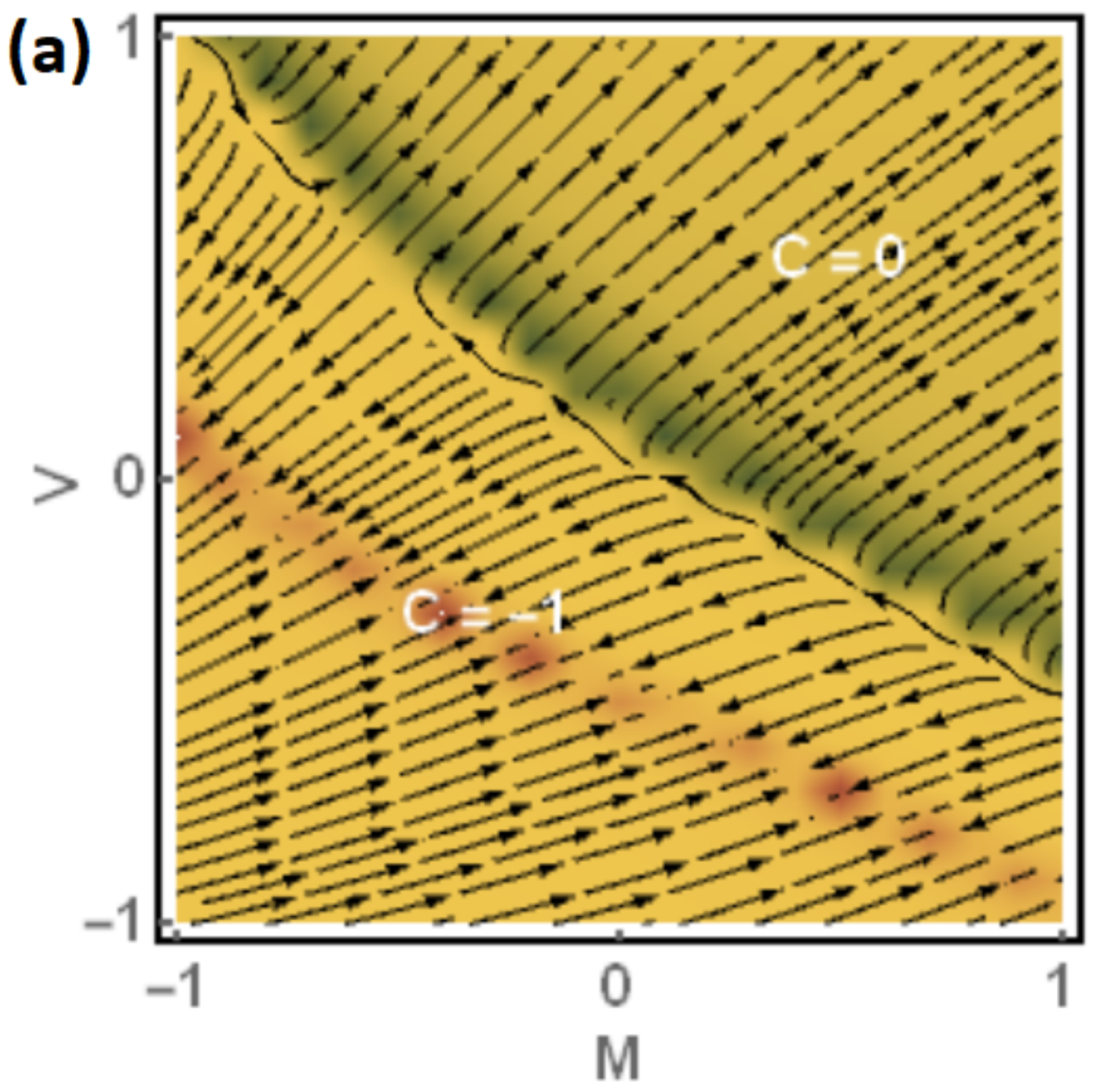}
\includegraphics[clip=true,width=0.99\columnwidth]{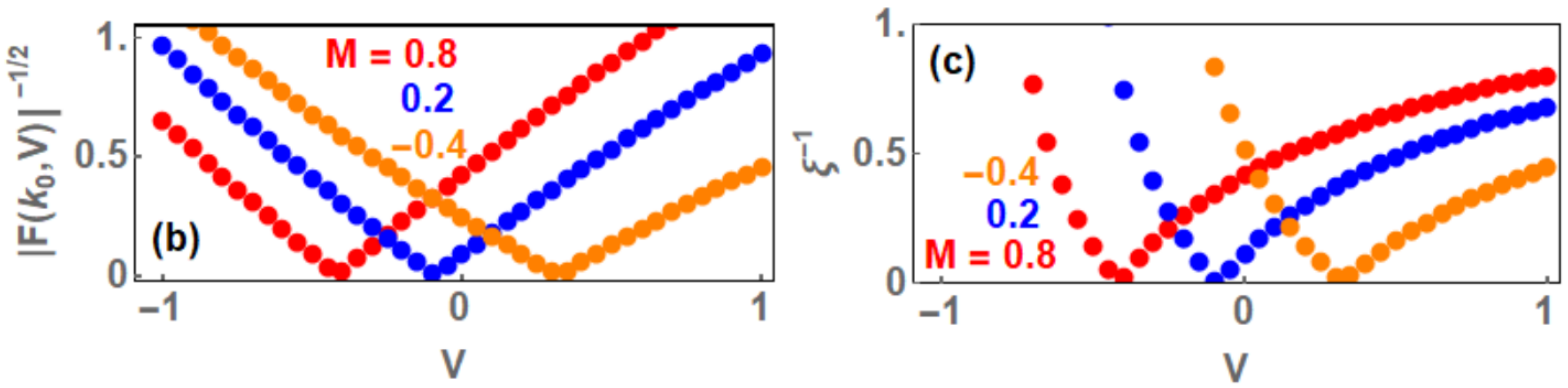}
\includegraphics[clip=true,width=0.99\columnwidth]{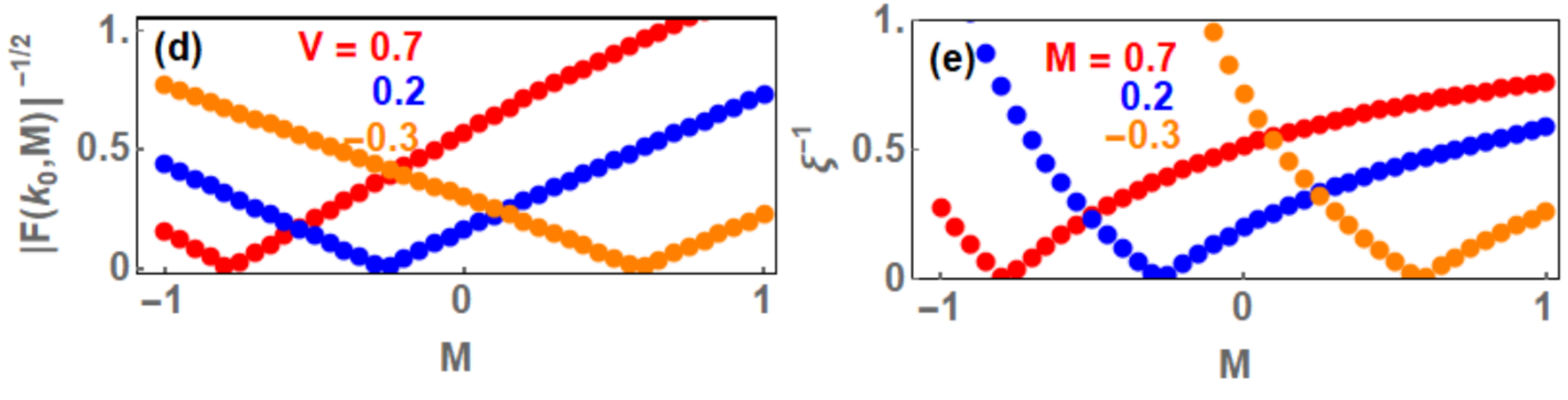}
\caption{ (a) The RG flow (black arrows) of 2D Chern insulator with density-density interaction in the ${\bf M}=(M,V)$ parameter space. The color scale indicates the high (green) and low (brown) flow rate on log scale $\ln|d{\bf M}/dl|$. The flow identifies a phase boundary (green line) between the topologically nontrivial ${\cal C}=-1$ and trivial ${\cal C}=0$ phase. The brown line indicates the fixed points in the topologically nontrivial phase. (b) and (c) show $|F({\bf k}_{0},{\bf M})|^{-1/2}$ and $\xi_{x}^{-1}$ along $V$ at several values of $M$, and (e) and (e) show them along $M$ at several values of $V$. The linear behavior near the critical point (where they vanish) indicate the critical exponents are $\gamma\approx 2$ and $\nu_{x}\approx 1$. } 
\label{fig:Chern_ee_RGflow}
\end{center}
\end{figure}

\subsection{2D Chern insulator with density-density interaction}

As a concrete example for 2D class A with frequency-independent self-energy, we examine the 2D Chern insulator, with the unperturbed Hamiltonian described by Eqs.~(\ref{general_2by2_Dirac_Hamiltonian}) and (\ref{2D_Chern_H0}), in the presence of the weak density-density interaction of the form of Eq.~(\ref{general_density_density_interaction}). 
We choose the $V_{\bf q}$ in Eq.~(\ref{general_density_density_interaction}) to have the form
\begin{eqnarray}
V_{\bf q}=V(2+\cos q_{x}+\cos q_{y})\;,
\end{eqnarray}
which may come from some nearest-neighbor density-density interaction between the two sublattices, by generalizing the argument in Sec.~\ref{sec:SSHnn} to 2D. Our aim is then to study the topological phase transition driven by the mass term in the unperturbed Hamiltonian and the nearest-neighbor interaction ${\bf M}=(M,V)$, which are treated as dimensionless parameters since the momentum and Fermi velocity are implicitly set to be dimensionless.

The CRG approach is applied to circumvent the 2D integral in Eq.~(\ref{2D_C_Fxy_dprime}). The curvature function $F({\bf k},{\bf M})$ is now only a function of momentum ${\bf k}=(k_{x},k_{y})$ since frequency has been integrated out. The RG flow of $M_{i}$ described by Eq.~(\ref{generic_RG_eq_discrete}) then requires to calculate the three mesh points $F({\bf k}_{0},{\bf M})$, $F({\bf k}_{0}+\Delta{\bf k},{\bf M})$, and $F({\bf k}_{0},{\bf M}+\Delta{\bf M}_{i})$, where $\Delta{\bf k}$ can be either $\Delta k_{x}{\hat{\bf k}}_{x}$ or $\Delta k_{y}{\hat{\bf k}}_{y}$. The result using ${\bf k}_{0}=(0,0)$ is shown in Fig.~\ref{fig:Chern_ee_RGflow}. Using the rule in Eq.~(\ref{identifying_Mc_Mf}), one identifies a phase boundary between the topologically nontrivial ${\cal C}=-1$ and trivial ${\cal C}=0$ phase in the ${\bf M}=(M,V)$ parameter space, which is a continuous line that passes through the well-known phase critical point in the noninteracting limit\cite{Bernevig13} ${\bf M}_{c}=(0,0)$ (slightly off in Fig.~\ref{fig:Chern_ee_RGflow} due to finite grid size). The fixed point in the nontrivial phase, at which the correlation length vanishes $\xi_{i}\rightarrow 0$, locates within the parameter space investigated. Figure \ref{fig:Chern_ee_RGflow} also shows the investigation of critical exponents defined in Eq.~(\ref{Fk0_xi_critical_exponent}). We extract $\gamma\approx 2$ and $\nu_{i}\approx 1$, same result as the noninteracting 2D Chern insulators and satisfy the scaling law $\gamma\approx \nu_{x}+\nu_{y}$\cite{Chen17}.

\section{Conclusions}

In summary, an RG approach is introduced into the Green's function formalism of weakly interacting TIs, as demonstrated particularly for 1D class BDI and 2D class A Dirac models under the influence of electron-electron and electron-phonon interactions. In the Green's function formalism, the topological invariant for these two symmetry classes is calculated from the integration of a curvature function calculated from the single-particle Green's functions. After the effect of interaction is included perturbatively into the Green's function, we show that the curvature function in these models diverges at the HSPs as the system approaches the topological phase transitions. We propose the CRG scheme based on this divergence, which is an iterative procedure to search for the trajectory in the parameter space along which the divergence reduces but the topological invariant remains unchanged, though which the topological phase transitions can be identified. The CRG scheme is proved to be a numerically efficient tool to investigate interacting models, since it circumvents the integration of topological invariant and only requires to calculate the curvature function on few points in the momentum or momentum-frequency space.

The divergence of curvature function also unveils a number of statistical aspects related to the quantum criticality in the Green's function formalism. The first is the correlation function introduced from the Fourier transform of the curvature function, which is generally a product of real space or spacetime Green's functions, their inverse, and the distance they travel over some segments. The correlation function decays exponentially due to the Ornstein-Zernike form of the curvature function. The divergence of curvature function further indicates that the correlation length or correlation time diverges at the topological phase transition, from which the scale invariance is interpreted. The critical exponents of the correlation length and correlation time, as well as that of the curvature function, are found to be constrained by a scaling law due to the conservation of topological invariant, and we find that interactions may or may not change the critical exponents. These results suggest that investigating the divergence of curvature function is a numerically efficient way to identify and characterize the topological phase transitions driven by either interacting or noninteracting parameters. We anticipate that the principles revealed in the article may also be applicable to the Green's function formalism of other symmetry classes, such as time-reversal invariant models with interactions\cite{Budich13,Lang13,Meng14}, which will be subject to further investigations.

\section{Acknowledgement}

The author acknowledges fruitful discussions with M. Sigrist, A. Schnyder, T. C. Lang, and R. Chitra.

\appendix

\section{Calculation of self-energy\label{appendix:self_energy}}

\begin{figure}[ht]
\begin{center}
\includegraphics[clip=true,width=0.99\columnwidth]{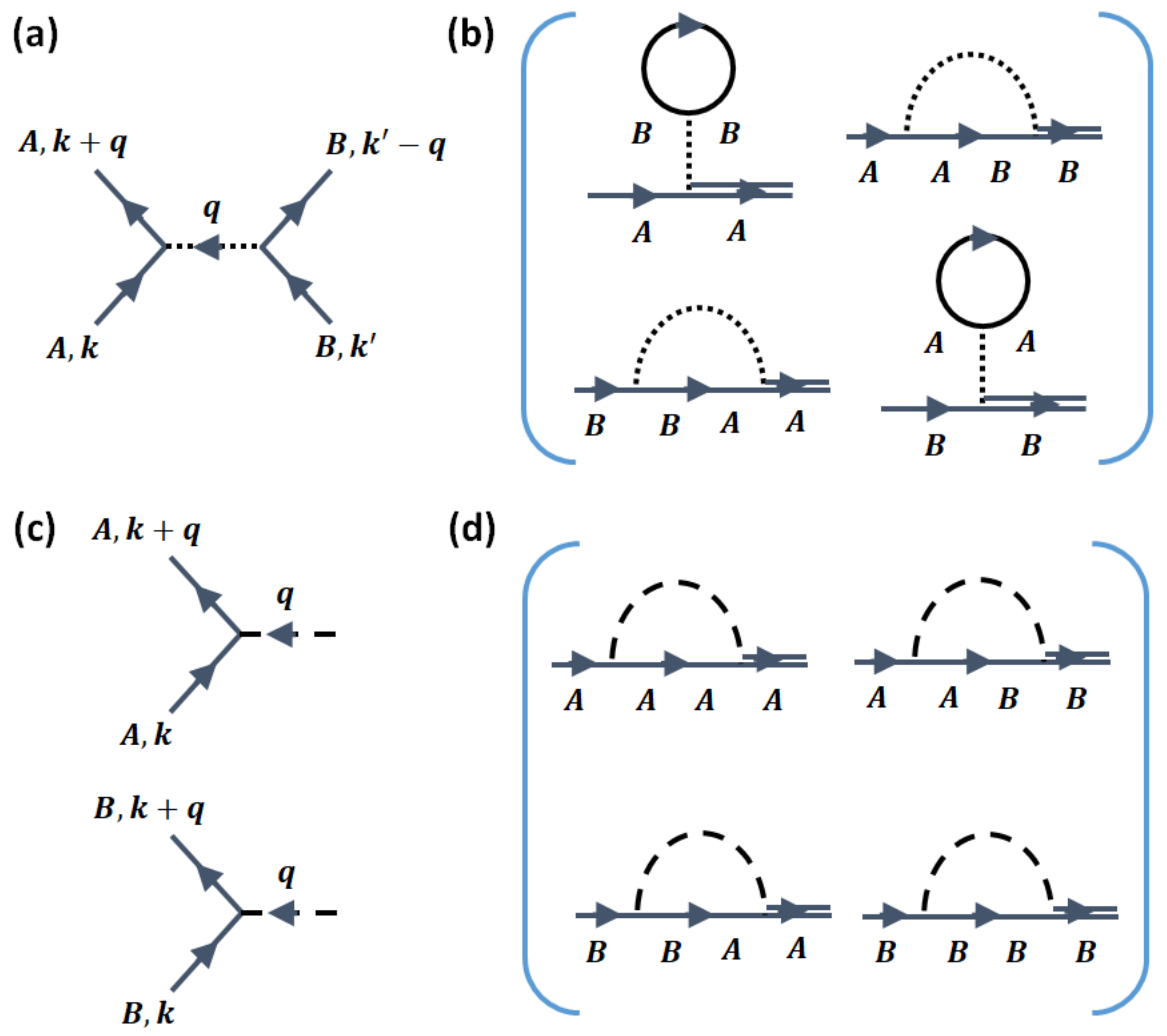}
\caption{ (a) The vertex for density-density interaction between sublattice $A$ and $B$ considered in this article. (b) The self-energy matrix for the density-density interaction case calculated up to one-loop Hartree-Fock level. (c) The vertex for electron-phonon interaction on each sublattice. (d) The self-energy matrix for the electron-phonon interaction case calculated up to one-loop. In (b) and (d), the four diagrams correspond to $\Sigma_{AA}$, $\Sigma_{AB}$, $\Sigma_{BA}$, and $\Sigma_{BB}$, respectively. } 
\label{fig:self_energy_Feymann_diagram}
\end{center}
\end{figure}

Here we detail the self-energy calculation for a $2\times 2$ Dirac model in the presence of the two kinds of interactions under consideration. First we consider the density-density interaction ${\cal H}={\cal H}_{0}+{\cal H}_{e-e}$ described in Sec.~\ref{sec:Green_fn_Dirac_interaction}. 
The interacting part of the Dyson'e equation in Eq.~(\ref{Dyson_equation}) up to one-loop level is
\begin{eqnarray}
&&\left(G_{0}\Sigma G\right)_{IJ}=\int_{0}^{\beta}d\tau_{1}\sum_{\bf pp^{\prime}q}V_{\bf q}
\nonumber \\
&&\times\langle T_{\tau}c_{I{\bf k}}(\tau)c_{A{\bf p+q}}^{\dag}(\tau_{1})c_{B{\bf p^{\prime}-q}}^{\dag}(\tau_{1})
c_{B{\bf p^{\prime}}}(\tau_{1})c_{A{\bf p}}(\tau_{1})c_{J{\bf k}}^{\dag}(0)\rangle.
\nonumber \\
\label{GSG_one_loop}
\end{eqnarray}
After a Fourier transform, the resulting self-energies, shown diagramatically in Fig.~\ref{fig:self_energy_Feymann_diagram}, are independent from the frequency $\Sigma_{IJ}({\bf k},i\omega_{n})=\Sigma_{IJ}({\bf k})$, and take the form
\begin{eqnarray}
\Sigma_{AA}({\bf k})&=&\sum_{\bf p}V_{\bf q=0}G_{0BB}({\bf p},\tau=0)\;,
\nonumber \\
\Sigma_{AB}({\bf k})&=&-\sum_{\bf q}V_{\bf q}G_{0AB}({\bf k+q},\tau=0)\;,
\nonumber \\
\Sigma_{BA}({\bf k})&=&-\sum_{\bf q}V_{\bf q}G_{0BA}({\bf k+q},\tau=0)\;,
\nonumber \\
\Sigma_{BB}({\bf k})&=&\sum_{\bf p}V_{\bf q=0}G_{0AA}({\bf p},\tau=0)\;.
\label{Hartree_Fock_self_energy}
\end{eqnarray}
Denoting $Q_{\bf k}=d_{1{\bf k}}-id_{2{\bf k}}$, using the convention that noninteracting Hamiltonian has $d_{0}=0$, and after a frequency sum, the unperturbed Green's function in the integrand of self-energy is
\begin{eqnarray}
&&G_{0AA}({\bf p},\tau=0)=\frac{1}{\beta}\sum_{\omega_{n}}G_{0AA}({\bf p},i\omega_{n})
\nonumber \\
&&=\frac{1}{2\beta}\sum_{\omega_{n}}\left[\frac{1+d_{3{\bf p}}/d_{\bf p}}{i\omega_{n}-d_{\bf p}}+\frac{1-d_{3{\bf p}}/d_{\bf p}}{i\omega_{n}+d_{\bf p}}\right]=\frac{1-d_{3{\bf p}}/d_{\bf p}}{2}\;,
\nonumber \\
&&G_{0BB}({\bf p},\tau=0)=\frac{1}{\beta}\sum_{\omega_{n}}G_{0BB}({\bf p},i\omega_{n})
\nonumber \\
&&=\frac{1}{2\beta}\sum_{\omega_{n}}\left[\frac{1-d_{3{\bf p}}/d_{\bf p}}{i\omega_{n}-d_{\bf p}}+\frac{1+d_{3{\bf p}}/d_{\bf p}}{i\omega_{n}+d_{\bf p}}\right]=\frac{1+d_{3{\bf p}}/d_{\bf p}}{2}\;,
\nonumber \\
&&G_{0AB}({\bf p+q},\tau=0)=\frac{1}{\beta}\sum_{\omega_{n}}G_{0AB}({\bf p+q},i\omega_{n})
\nonumber \\
&&=\frac{e^{-i\alpha_{\bf p+q}}}{2\beta}\sum_{\omega_{n}}\left[\frac{|Q_{\bf p+q}|/d_{\bf p+q}}{i\omega_{n}-d_{\bf p+q}}-\frac{|Q_{\bf p+q}|/d_{\bf p+q}}{i\omega_{n}+d_{\bf p+q}}\right]
\nonumber \\
&&=-\frac{d_{1{\bf p+q}}-id_{2{\bf p+q}}}{2d_{\bf p+q}}=G_{0BA}({\bf p+q},\tau=0)^{\ast}\;.
\end{eqnarray}
Consequently, the self-energy reads
\begin{eqnarray}
&&\Sigma_{AA}({\bf k})=V_{\bf q=0}\sum_{\bf p}\frac{1}{2}\left(1+\frac{d_{3{\bf p}}}{d_{\bf p}}\right)\;,
\nonumber \\
&&\Sigma_{BB}({\bf k})=V_{\bf q=0}\sum_{\bf p}\frac{1}{2}\left(1-\frac{d_{3{\bf p}}}{d_{\bf p}}\right)\;,
\nonumber \\
&&\Sigma_{AB}({\bf k})=\sum_{\bf q}V_{\bf q}\frac{d_{1{\bf k+q}}-id_{2{\bf k+q}}}{2d_{\bf k+q}}=\Sigma_{BA}({\bf k})^{\ast},\;\;\;
\end{eqnarray}
which is manifestly frequency-independent. It is well known that the Hartree term $\Sigma_{AA}=\Sigma_{BB}$ is a constant that effectively gives a chemical potential that contributes to the $d_{0}$ component according to Eq.~(\ref{renormalized_d_vector_mu}). Note that for 1D class BDI models one has $d_{3}=0$.

We proceed to consider the 2D class A Dirac model with electron-phonon interaction ${\cal H}={\cal H}_{0}+{\cal H}_{e-ph}$. The leading order expansion in the Dyson's equation is
\begin{eqnarray}
&&\left(G_{0}\Sigma G\right)_{IJ}=-\sum_{\bf qq^{\prime}}M_{\bf q}M_{\bf q^{\prime}}\int_{0}^{\beta}d\tau_{1}\int_{0}^{\beta}d\tau_{2}
\nonumber \\
&&\times\langle T_{\tau}A_{\bf q}(\tau_{1})A_{\bf q^{\prime}}(\tau_{2})\rangle
\nonumber \\
&&\times\sum_{\bf pp^{\prime}}\langle T_{\tau}c_{I{\bf k}}(\tau)\left[c_{A{\bf p+q}}^{\dag}(\tau_{1})c_{A{\bf p}}(\tau_{1})+c_{B{\bf p+q}}^{\dag}(\tau_{1})
c_{B{\bf p}}(\tau_{1})\right]
\nonumber \\
&&\times\left[c_{A{\bf p^{\prime}+q^{\prime}}}^{\dag}(\tau_{1})c_{A{\bf p^{\prime}}}(\tau_{1})+c_{B{\bf p^{\prime}+q^{\prime}}}^{\dag}(\tau_{1})
c_{B{\bf p^{\prime}}}(\tau_{1})\right]
c_{J{\bf k}}^{\dag}(0)\rangle.
\nonumber \\
\label{GSG_one_loop_eph}
\end{eqnarray}
After a Fourier transform, the matrix element of the self-energy with Matsubara frequency $i\omega_{n}$ reads 
\begin{eqnarray}
&&\Sigma_{IJ}(k,i\omega_{n})
\nonumber \\
&&=-\frac{1}{\beta}\sum_{\omega_{m}}\sum_{\bf q}M_{\bf q}^{2}D_{0}({\bf q},i\omega_{m})G_{0IJ}({\bf k-q},i\omega_{n}-i\omega_{m})\;.
\nonumber \\
\end{eqnarray}
Upon a frequency sum, the $\Sigma_{IJ}$ can be evaluated, which contains the Bose distribution function $N_{0}$ and the Fermi distribution function $n_{F}$ with appropriate energy variables. In the zero temperature limit $T\rightarrow 0$, $N_{0}=0$ and $n_{F}(x)=\theta(-x)$, the self-energy is 
\begin{eqnarray}
&&\Sigma_{AA}({\bf k},i\omega_{n})=\sum_{\bf q}M_{\bf q}^{2}
\nonumber \\
&&\times\left[\frac{(1+d_{3{\bf k-q}}/d_{\bf k-q})/2}{i\omega_{n}-\omega_{0}-d_{\bf k-q}}+\frac{(1-d_{3{\bf k-q}}/d_{\bf k-q})/2}{i\omega_{n}+\omega_{0}+d_{\bf k-q}}\right]\;,
\nonumber \\
&&\Sigma_{BB}({\bf k},i\omega_{n})=\sum_{\bf q}M_{\bf q}^{2}
\nonumber \\
&&\times\left[\frac{(1-d_{3{\bf k-q}}/d_{\bf k-q})/2}{i\omega_{n}-\omega_{0}-d_{\bf k-q}}+\frac{(1+d_{3{\bf k-q}}/d_{\bf k-q})/2}{i\omega_{n}+\omega_{0}+d_{\bf k-q}}\right]\;,
\nonumber \\
&&\Sigma_{AB}({\bf k},i\omega_{n})=\sum_{\bf q}M_{\bf q}^{2}\frac{Q_{\bf k-q}}{2d_{\bf k-q}}
\nonumber \\
&&\times\left[\frac{1}{i\omega_{n}-\omega_{0}-d_{\bf k-q}}-\frac{1}{i\omega_{n}+\omega_{0}+d_{\bf k-q}}\right]\;,
\nonumber \\
&&\Sigma_{BA}({\bf k},i\omega_{n})=\sum_{\bf q}M_{\bf q}^{2}\frac{Q_{\bf k-q}^{\ast}}{2d_{\bf k-q}}
\nonumber \\
&&\times\left[\frac{1}{i\omega_{n}-\omega_{0}-d_{\bf k-q}}-\frac{1}{i\omega_{n}+\omega_{0}+d_{\bf k-q}}\right]\;,
\end{eqnarray} 
The calculation of derivative of self-energy $\partial_{i}\Sigma_{IJ}$ is then straight forward, although the expressions are rather lengthy and we omit here. At a given $({\bf k},i\omega_{n})$, these matrix elements satisfy
\begin{eqnarray}
\Sigma_{AA}=-\Sigma_{BB}^{\ast}\;,\;\;\;\Sigma_{AB}=\Sigma_{BA}^{\ast}\;,
\end{eqnarray}
and so are their derivatives, so numerically one only has to calculate $\left\{\Sigma_{AA},\Sigma_{AB}\right\}$ and their derivatives $\left\{\partial_{i}\Sigma_{AA},\partial_{i}\Sigma_{AB}\right\}$. We then take the discrete Matsubara frequency continuously $i\omega_{n}\rightarrow i\omega$ in $\Sigma_{IJ}$ and $\partial_{i}\Sigma_{IJ}$ to calculate the self-energy-renormalized ${\bf d}^{\prime}$-vector according to Eq.~(\ref{renormalized_d_vector_mu}), and then obtain the curvature function according to Eq.~(\ref{TrGGGGGG_to_dprime_to_FkM}).

\bibliography{Literatur}

\end{document}